\documentclass[twoside]{article}
\usepackage{color,graphicx,lscape,here,geometry,setspace,multirow,enumitem}
\usepackage[utf8]{inputenc}  
\usepackage[natbib=true,sorting=nyt,uniquelist=false,maxcitenames=2,
mincitenames=1,backend=biber,style=apa,useprefix=false, uniquename=false]{biblatex}
\usepackage[dvipsnames]{xcolor}
\usepackage{graphicx}
\usepackage{authblk}
\usepackage{silence}
\usepackage{longtable}
\usepackage{bbm}
\usepackage[ruled,vlined]{algorithm2e}
\usepackage{rotating}
\usepackage{adjustbox}
\usepackage{multirow}
\usepackage{gensymb}
\usepackage{enumitem}

\usepackage{csquotes}

\usepackage{booktabs}
\usepackage{threeparttable}

\usepackage{amsmath}
\usepackage{amssymb}
\usepackage{algorithm2e}

\definecolor{nblue}{RGB}{87, 16, 110}
\usepackage[colorlinks=true,urlcolor=blue,linkcolor=blue,citecolor=blue]{hyperref}
\usepackage{bigstrut}

\usepackage{tabularx}
\usepackage{dcolumn}

\usepackage{etoolbox} 
\makeatletter
\patchcmd{\BR@backref}{\newblock}{\newblock[}{}{}
\patchcmd{\BR@backref}{\par}{]\par}{}{}
\makeatother

\usepackage{pdflscape}
\usepackage{afterpage}
\usepackage{rotate}
\usepackage{longtable}
\usepackage{array}
\newcolumntype{C}[1]{>{\centering\arraybackslash}p{#1}}

\usepackage{comment}
\usepackage{mathtools}

\usepackage[hang,flushmargin]{footmisc} 
\usepackage[british]{babel}
\usepackage{multicol}
\usepackage{amsmath, amssymb}
\pdfoutput=1
\usepackage[title,titletoc]{appendix}
\makeatletter

\renewenvironment{appendices}{%
    \begin{oldappendices}%
    \renewcommand{\thefigure}{\ifnum \c@section>\z@ \thesection.\fi\@arabic\c@figure}%
    \@addtoreset{figure}{section}%
    \renewcommand{\thetable}{\ifnum \c@section>\z@ \thesection.\fi\@arabic\c@table}%
    \@addtoreset{table}{section}}{%
    \end{oldappendices}%
}\makeatother

\usepackage{titlesec} 
\titleformat{\section}[block]{\large}{\thesection. }{0em}{\MakeUppercase} 
\titleformat{\subsection}[block]{\large}{\thesubsection. }{0em}{\itshape} 
\titleformat{\subsubsection}[block]{\large}{}{0em}{\itshape} 

\usepackage{sectsty}
\allsectionsfont{\normalsize}

\makeatletter

\@addtoreset{equation}{section}
\makeatother
\usepackage{bm}
\usepackage[font=small]{caption}
\usepackage{subcaption}

\makeatletter\let\p@subfigure\thefigure\makeatother

\usepackage{fancyhdr} 
\usepackage[capitalise]{cleveref}
\crefname{chapter}{Chapter}{Chapters}
\crefname{section}{Section}{Sections}
\crefname{subsection}{Section}{Sections}
\crefname{subsubsection}{Section}{Sections}
\crefname{figure}{Figure}{Figures}
\crefname{table}{Table}{Tables}
\crefname{equation}{Equation}{Equations}
\crefname{appendix}{Appendix}{Appendices}
\crefname{appendices}{Appendix}{Appendices}
\crefname{algocf}{alg.}{algs.}
\Crefname{algocf}{Algorithm}{Algorithms}

\crefname{appsec}{Appendix}{Appendices}

\renewcommand\labelenumi{(\roman{enumi})}
\renewcommand\theenumi\labelenumi

\title{
\vspace{-1cm}
\LARGE{Flexible Bayesian Modelling of Age-Specific Counts in Many Demographic Subpopulations}\thanks{Correspondence to Gregor Zens, International Institute for Applied Systems Analysis. Schlossplatz 1, 2361 Laxenburg, Austria. E-mail: \texttt{\href{mailto:zens@iiasa.ac.at}{zens@iiasa.ac.at}}. An earlier version of this paper circulated under the title `Bayesian Modeling of Demographic Data in Many Subpopulations'.}}
\author{\large{\uppercase{Gregor Zens}}\\
\vspace*{0em}
\normalsize{\textit{International Institute for Applied Systems Analysis (IIASA)}\\\textit{Wittgenstein Centre for Demography and Global Human Capital (WIC)}}\\
\ \\
\today
\vspace{-1cm}
}

\date{}


\def\equationautorefname~#1\null{%
  Eq.~(#1)\null
}
\def\equationautorefname~#1\null{
Eq.~(#1)\null
}

\makeatletter
\renewcommand\paragraph{\@startsection{paragraph}{4}{\z@}%
          {-3.25ex\@plus -1ex \@minus -.2ex}%
          {1.5ex \@plus .2ex}%
          {\normalfont\itshape}}
\makeatother

\begin{document}

\maketitle

\begin{abstract} 

Analysing age-specific mortality, fertility, and migration patterns is a crucial task in demography with significant policy relevance. In practice, such analysis is challenging when studying a large number of subpopulations, due to small observation counts within groups and increasing demographic heterogeneity between groups. This article proposes a Bayesian model for the joint analysis of age-specific counts in many, potentially small, demographic subpopulations. The model utilizes smooth latent factors to capture common age-specific patterns across subpopulations and facilitates additional information sharing through a hierarchical prior. It provides smoothed estimates of the latent age pattern in each subpopulation, allows testing for heterogeneity, and can be used to assess the impact of covariates on the demographic process. An in-depth case study of age-specific immigration flows to Austria, disaggregated by sex and 155 countries of origin, is discussed. Comparative analysis demonstrates that the model outperforms commonly used benchmark frameworks in both in-sample imputation and out-of-sample predictive exercises.

\end{abstract}


\newpage

\renewcommand{\thepage}{\arabic{page}}

\newrefsection

\section{Introduction}

Analysing data recorded by age and additional individual characteristics, such as sex and education, plays a prominent role in demographic research. Modelling such multidimensional data is crucial for informing policymakers and population projections, for exploring the drivers of heterogeneity among subpopulations, and for predicting the demographic characteristics of populations for which data are unavailable. The latter is a fundamental challenge, particularly in the context of developing countries and when conducting subnational analyses. 

\begin{figure}
     \centering
     \includegraphics[width=\textwidth]{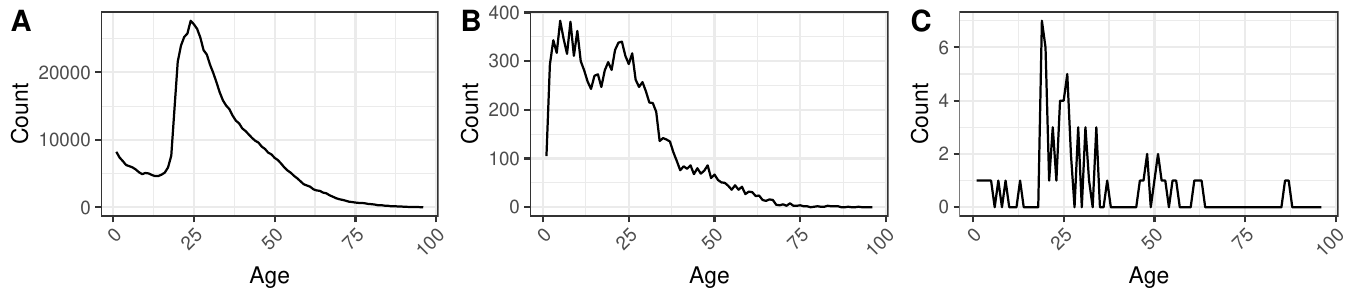}
        \caption{Age-specific immigration flows to Austria, 2016-2020. $x$-axis shows age and $y$-axis shows counts. Panel (A) shows the aggregate counts. Panel (B) shows counts of male immigrants from Syria, characterized by patterns that differ substantially from the aggregate. Panel (C) shows counts of female immigrants from Guatemala, with substantial stochastic variation.}
        \label{fig:motivating}
\end{figure}

In practice, modelling multidimensional demographic data can be difficult. This is particularly true when considering many, potentially small subpopulations. To illustrate this, consider the data shown in Fig.~\ref{fig:motivating}. Each of the three panels shows immigrant flows as a function of age. These migrant counts are based on register data from Austria and include all international immigrants who arrived in the country between 2016 and 2020. Details on the data set are provided in Section~\ref{sec:application}. Panel (A) shows the aggregate age-specific immigrant counts, revealing typical regularities of international migration intensity across the life course, such as the majority of migration events taking place between ages 20 and 40. Panel (B) shows immigration flows by age for male immigrants from Syria. This subpopulation deviates from the aggregate pattern, with a much larger proportion of immigrants falling into the younger age groups. This reflects the distinct refugee dynamics in this subpopulation. Panel (C) shows immigrant counts by age for female immigrants from Guatemala. Only a small number of migrants are members of this subpopulation, resulting in significant stochastic variation. This makes it difficult to make reliable statements about the underlying age pattern on the basis of the raw data alone, and motivates the use of statistical models to analyse the generating demographic process. 


Model-based analysis of demographic phenomena in such multipopulation settings is, however, challenging, precisely due to the interplay of small observation counts and demographic heterogeneity. Models need to be flexible enough to accurately capture systematic heterogeneity between subpopulations. At the same time, models must be robust to avoid misinterpreting noisy patterns in small subpopulations as systematic demographic information. Additionally, handling the complex interdependencies and correlations among subpopulations in high-dimensional settings demands parsimonious, yet comprehensive modelling approaches.

To address these challenges, this article presents a Bayesian model for the demographic analysis of age-specific counts in many, potentially small, subpopulations. The model is based on three widely recognized assumptions about demographic processes, which are formalized and combined into a single hierarchical framework. Its key applications include producing smoothed estimates of age patterns for each subpopulation, testing for heterogeneity across subpopulations, exploring systematic drivers of this heterogeneity, producing predictions, as well as probabilistic uncertainty quantification. These are illustrated through an in-depth analysis of data on international migration flows to Austria, disaggregated by age, sex, and country of origin, resulting in a total of 300 subpopulations. The results indicate pronounced heterogeneity in age patterns across subpopulations, with conflict in origin countries identified as one of the key modulating factors. Systematic simulation studies and real data exercises further demonstrate the relative advantage of the model over competing models for in-sample smoothing, imputation of partially missing data, and out-of-sample forecasting tasks.

The rest of this article is structured as follows. Section~\ref{sec:literature} summarizes related literature. Section~\ref{sec:statistics} provides the statistical details of the proposed framework. Section~\ref{sec:application} applies the framework to data on international immigration to Austria. Section~\ref{sec:simulations} provides insights into the comparative performance of the model based on simulated and real data. Section~\ref{sec:conclusion} concludes and provides directions for future research.

\section{Related Literature}
\label{sec:literature}

Quantitative methods have long been integral to understanding population structures, fertility, mortality, and migration. Bayesian demography has recently emerged as a dynamic subfield (\citealp{raftery2014bayesian}; \citealp{shang2016multilevel}; \citealp{bijak2016bayesian}; \citealp{bryant2018bayesian}; \citealp{alexander2024developing}). Specialized Bayesian methods have been developed for the analysis of mortality (\citealp{dellaportas2001bayesian}; \citealp{czado2005bayesian}; \citealp{raftery2014joint}; \citealp{antonio2015bayesian};  \citealp{alexander2017flexible}; \citealp{schmertmann2018bayesian};  \citealp{fung2019cohort}; \citealp{alexopoulos2019bayesian}; \citealp{wang2023multi}; \citealp{dharamshi2025jointly}), fertility (\citealp{assuncao2002bayesian}; \citealp{schmertmann2014bayesian}; \citealp{castro2015regional}; \citealp{schmertmann2019bayesian}; \citealp{ellison2020forecasting}) and migration patterns (\citealp{bijak2008bayesian}; \citealp{bijak2010forecasting}; \citealp{raymer2013integrated}; \citealp{wisniowski2014forecasting}; \citealp{azose2015bayesian}; \citealp{wisniowski2016integrated}; \citealp{zhang2020bayesian}; \citealp{rampazzo2021framework}; \citealp{bijak2022towards}; \citealp{bijak2024uncertainty}), complementing and extending traditional demographic modelling tools.

In this broader context, the main contribution of the present article is to extend the literature on statistical analysis of age-specific patterns in multipopulation demographic data. Numerous statistical models have been developed for this task. The remainder of this section summarizes three key ideas and underlying assumptions about demographic processes that are regularly put forward in this literature. These will later be formalized and combined into a single unified hierarchical model.

\paragraph{Assumption \textsc{AS1}: Latent Commonalities}

A frequent assumption is the existence of common underlying patterns shared by many or all subpopulations in the data. This idea arises naturally from the regularity of demographic processes in different contexts. Examples include a higher-than-average probability of dying in old age or a higher-than-average probability of migrating between the ages of 20 and 40. Due to such regularities, methods based on principal component analysis (PCA) and singular value decompositions (SVDs) are highly popular in statistical analysis of demographic data, see for instance \citet{alexander2017flexible}, \citet{clark2019general}, and \citet{dharamshi2025jointly} who focus on latent commonalities in the age dimension, or the Lee-Carter framework and its extensions (\citealp{lee1992modeling}; \citealp{li2005coherent}; \citealp{shang2011point}; \citealp{wisniowski2015bayesian}) focusing on latent commonalities in the time dimension. 

Traditional demographic methods based on fitting pre-specified parametric functions to demographic data can also be seen as exploiting the concept of common underlying patterns. For example, the Rogers-Castro migration model (\citealp{rogers1978model}) or the Heligman-Pollard mortality model (\citealp{heligman1980age}) postulate that age-specific demographic patterns in different populations can typically be reconstructed by appropriately weighting and combining (pre-specified) common underlying components. Finally, similar ideas about latent commonalities underlie mixture models for demographic data (\citealp{mazzuco2018mortality}) as well as the TOPALS framework (\citealp{de2011new}; \citealp{de2012smoothing}).

\paragraph{Assumption \textsc{AS2}: Smoothing Across Ages}

A common strategy in demographic modelling is to assume that demographic processes are smooth in the age dimension. This translates into the assumption that, for example, the demographic characteristics and behaviours of 5-year-olds and 6-year-olds are expected to be relatively more similar than the characteristics of 5-year-olds and 60-year-olds a priori. This assumption motivates methods that make direct use of tools such as spline smoothing, kernel smoothing, or related ideas from functional data analysis. Such approaches have been extensively studied and applied in demographic literature; see for example \citet{mcneil1977spline}, \citet{hyndman2007robust}, \citet{de2011new}, \citet{de2012smoothing}, \citet{camarda2012mortalitysmooth}, \citet{hyndman2013coherent}, \citet{gonzaga2016estimating}, \citet{dyrting2020smoothing}, or \citet{pavone2024learning}.

\paragraph{Assumption \textsc{AS3}: Information Sharing Among Similar Subpopulations}

A third key idea is based on the assumption that some notion of `similarity' can be formulated across subpopulations and that similar subgroups will be characterized by similar demographic processes. In terms of the motivating application of Austrian immigration patterns, this assumption might imply, for example, that male migrants from Kenya and Ethiopia are a priori expected to be characterized by a relatively more similar age pattern than male migrants from Kenya and female migrants from Germany. Different notions of similarity have been proposed, based on spatial proximity (\citealp{alexander2017flexible}), temporal proximity (\citealp{lee1992modeling}; \citealp{hyndman2007robust}), or based on externally available covariates (\citealp{clark2019general}). Statistically, the idea of sharing information across subpopulations based on similarity is most often formalized using hierarchical modelling frameworks (\citealp{susmann2022temporal}). Hierarchical models allow demographic processes to be reconstructed, even in subpopulations with sparse data, by borrowing information from `similar' subpopulations. In addition, hierarchical structures facilitate forecasting, imputation of missing data, and projection exercises.

\paragraph{A Unified Perspective}

While these three assumptions are ubiquitous in empirical demography, their application in isolation suffers from several shortcomings. For example, when dealing with noisy multipopulation data, applying flexible demographic methods such as smoothing splines separately to each subpopulation -- without sharing information about `typical' patterns across groups -- can easily lead to overfitting noisy, idiosyncratic patterns within subpopulations. In contrast, traditional demographic models often rely on predefined parametric functions that are explicitly designed to mimic commonly shared empirical patterns, and thus have the potential to avoid such within-group overfitting. However, these traditional models often lack the flexibility needed to capture `atypical', but genuine demographic heterogeneity that may be present in detailed multipopulation data. More recent methods based on extracting latent commonalities using SVD or PCA can strike a balance between flexibility and robustness, but have significant potential for error propagation when facing multipopulation data. This is because if the data contains small and noisy subpopulations, the extracted principal components will contain a certain amount of noise as well. This noise is then multiplicatively reintroduced into the PCA-based predictions for each subpopulation. Empirical illustrations of these phenomena are given in Section~\ref{sec:simulations}.

This paper proposes a modelling framework for multidimensional demographic analysis that superimposes all three assumptions jointly in a single probabilistic framework. The framework exploits the advantages of the three stated assumptions without the shortcomings of applying them in isolation. The model represents potentially noisy and heterogeneous multipopulation data using a small number of latent smooth components. The relative importance of these components for a given subpopulation is based on the similarity between subpopulations. This approach allows to efficiently handle a large number of subpopulations. Employing a Bayesian setup for the model further leads to fully probabilistic inference and uncertainty quantification.

From a statistical perspective, the proposed framework is a Bayesian functional probabilistic PCA model for count data. The loadings are modelled using a hierarchical regression structure, allowing for information sharing across subpopulations. The statistical approach thus integrates ideas from Bayesian factor models (\citealp{conti2014bayesian}; \citealp{fruhwirth2024sparse}), Bayesian probabilistic PCA for count outcomes (\citealp{chiquet2018variational}), related Bayesian demographic models  (\citealp{czado2005bayesian}; \citealp{alexander2017flexible}), functional data analysis (\citealp{hyndman2007robust}; \citealp{montagna2012bayesian}; \citealp{kowal2020bayesian}), and literature on Bayesian smoothing and shrinkage priors (\citealp{lang2004bayesian}; \citealp{pii-veh:spa}).

\section{Statistical Framework}
\label{sec:statistics}

Let $y_{i,x}$ denote a demographic count outcome observed at a discrete age $x$ in a subpopulation $i$.\footnote{For notational simplicity, a single subscript is used to refer to subpopulations, rather than a panel-type notation based on multiple subscripts.} The counts $y_{i,x}$ are modelled as observations from a Poisson distribution\footnote{While the focus on a count data framework is motivated by its application in empirical demography, (\ref{eq:eq1}) can be easily replaced by alternative likelihood specifications, for example for Gaussian outcomes or for binary and binomial outcomes based on data augmentation techniques (\citealp{tan-won:cal}; \citealp{pol-sco:dat}; \citealp{zens2023ultimate}).}

\begin{equation}
\label{eq:eq1}
{y}_{i,x} \sim \mathcal{P}(e^{z_{i,x}}).
\end{equation}

The log-mean parameter $z_{i,x}$ is a noisy realization of a Gaussian model

\begin{equation}
\label{eq:eq_z}
{z}_{i,x} = \alpha_i + z_i(x) + O_{i,x} + \varepsilon_{i,x} \quad \quad \varepsilon_{i,x} \sim \mathcal{N}(0, \sigma^2),
\end{equation}

where $\alpha_i$ is a subpopulation-specific intercept, ${z}_{i}(x)$ is a smooth function of age and $\varepsilon_{i,x}$ is a subpopulation and age-specific iid zero mean Gaussian noise term with variance $\sigma^2$. $O_{i,x}$ is a known offset or exposure term, such as the logarithm of the total population count for subpopulation $i$ and age $x$. Known exposure terms are typically used in studies of mortality, fertility, or emigration, where a `population at risk' can be clearly defined and the outcome of interest is usually a rate. When modelling immigration patterns, there is no clear definition of an exposed population and usually $O_{i,x}=0$ is fixed. The Poisson lognormal model (\citealp{aitchison1989multivariate}) defined by (\ref{eq:eq1}) and (\ref{eq:eq_z}) effectively handles zero counts, accounts for overdispersion and hence helps to adequately quantify estimation uncertainty. 

It is further assumed that the functions ${z}_{i}(x)$ can be written as a linear combination of $Q$ $(q=1,\dots,Q)$ unknown, smooth zero-mean functions $\Phi_q(x)$, weighted by subpopulation specific loadings $\lambda_{i,q}$ such that 

\begin{equation}
\label{eq:z_factors}
{z}_{i}(x) = \sum_{q=1}^Q \Phi_q(x) \lambda_{i,q}.
\end{equation}

 The latent functions $\Phi_q(x)$ represent underlying commonalities across subpopulations, reflecting assumption \texttt{AS1} (see Section~\ref{sec:literature}). The loadings $\lambda_{i,q}$ determine how important a particular function $\Phi_q(x)$ is in describing the observed patterns within a given subpopulation $i$. Assuming that $Q << N$ allows for an adequate description of the data, this approach is related to probabilistic principal component models for counts (\citealp{chiquet2018variational}), functional principal component frameworks (\citealp{ramsay2005principal}), functional regression models (\citealp{kowal2020bayesian}), and functional factor models (\citealp{montagna2012bayesian}). In related demographic literature, the functions $\Phi_q(x)$ are typically estimated via an SVD of the log counts or log rates (\citealp{alexander2017flexible}; \citealp{clark2019general}; \citealp{dharamshi2025jointly}). In this paper, we treat them explicitly as parameters that are surrounded by uncertainty and that need to be estimated along with the rest of the model parameters. This approach has several advantages, including an appropriate quantification of uncertainty, straightforward estimation from partially incomplete data, and a lower likelihood of capturing noise in the factors, due to the presence of an error term in the model. To facilitate estimation, we assume that the unknown functions $\Phi(x)$ can be represented as a linear combination of $K (k=1, \dots, K)$ B-spline basis functions $\bm{B}_{k}(x)$ 

\begin{equation}
\label{eq:splineapprox}
\Phi_q(x) = \sum_{k=1}^K f_{q,k} \bm{B}_{k}(x)
\end{equation}

where $f_{q,k}$ is the spline base coefficient of knot $k$ and function $q$. A Bayesian approach to estimation is pursued to allow for probabilistic inference. The Bayesian paradigm requires the choice of appropriate prior distributions for all model parameters. The prior distributions are chosen in an informative way that reflects assumptions on smoothing (\texttt{AS2}) and information sharing (\texttt{AS3}); see Section~\ref{sec:literature}. 

\subsection{Smoothing Mechanism}

It is assumed that the underlying latent functions $\Phi_q(x)$ are a priori smooth, in the sense that small changes in $x$ imply only small changes in $\Phi_q(x)$.  Formally, a smoothing prior on the spline coefficients $f_{q,k}$ is used to introduce this assumption. In particular, we follow \citet{lang2004bayesian} and work under a stochastic difference penalty in the form of a second-order random walk on $f_{q,k}$, such that, a priori

\begin{equation}
\label{eq:smoothing1}
f_{q,k} = 2f_{q,k-1} - f_{q,k-2} + u_{q,k}
\end{equation}

where $u_{q,k}$ is a noise term. Flat priors are specified for the initial values $f_{q,1}$ and $f_{q,2}$ for each $q$. To allow for locally adaptive smoothing behaviour, we let 

\begin{equation}
\label{eq:smoothing2}
u_{q,k} \sim \mathcal{N}\left(0, \frac{\tau_q}{\kappa_{q,k}}\right)
\end{equation}

with $\kappa_{q,k} \sim \mathcal{G}(\frac{1}{2},\frac{1}{2})$. Marginally, this implies a Cauchy distribution for $u_{q,k}$. The prior specification (\ref{eq:smoothing1}) - (\ref{eq:smoothing2}) regularizes `neighbouring' spline coefficients to be similar, resulting in smoother estimates of the underlying components $\Phi_q(x)$. The assumption underlying this prior choice is that the underlying age patterns change gradually, with the aim of making the estimates $\Phi_q(x)$ more robust to random fluctuations that may be present in small subpopulations. In addition, this prior renders the analysis less sensitive to knot placement (\citealp{lang2004bayesian}).

\subsection{Information Sharing}

It is assumed that `similar' subpopulations are a priori characterized by `similar' demographic processes. Formally, this is implemented by assuming a hierarchical structure for the loadings $\lambda_{i,q}$ and intercepts $\alpha_i$ which determine the shape and level of the demographic process, respectively. In demographic trend analysis, such hierarchical structures often involve time series elements (\citealp{lee1992modeling}; \citealp{hyndman2007robust}; \citealp{susmann2022temporal}). In subnational mortality modelling, \citet{alexander2017flexible} allow for similarity based on spatial proximity of subnational units. \citet{dharamshi2025jointly} present a more complex approach, based on the geographic nesting structure of subnational units. We assume that the loadings $\lambda_{i,q}$ and intercepts $\alpha_i$ are a function of observed subpopulation-specific covariates:

 \begin{equation}
 \label{eq:weights}
 \begin{split}
 \alpha_i & \sim \mathcal{N}(\bm{w}_i'\bm{\delta}, \sigma^2_\alpha)\\
 \lambda_{i,q} &\sim \mathcal{N}(\bm{w}_i'\bm{\beta}_q, \sigma^2_{\lambda,q}).
 \end{split}
 \end{equation}

 where $\bm{w}_i$ is a $R \times 1$ vector of observed covariates describing subpopulation $i$, $\bm{\beta}_q$ and $\bm{\delta}$ are $R \times 1$ vectors of linear regression coefficients, and $\sigma^2_{\lambda,q}$ and $\sigma^2_\alpha$ are variance terms.\footnote{Extending the model to include, e.g., a temporal smoothing component is straightforward. This is not considered in this paper, partially due to the challenges of modelling temporal patterns in migration data (\citealp{bijak2019assessing}).} This specification allows for heterogeneity around the conditional prior means of the loadings $\lambda_{i,q}$ and intercepts $\alpha_i$. As will be discussed in Section~\ref{sec:covariates}, covariate effects on the shape and level of demographic processes can be derived from (\ref{eq:weights}) to explore drivers of heterogeneity across subpopulations. In Section~\ref{sec:scenario}, it is demonstrated that (\ref{eq:weights}) is also useful for probabilistic projection exercises.
 
The hierarchical information sharing mechanism (\ref{eq:weights}) allows subpopulations with informative data to contribute to demographic process estimates in `similar' subpopulations, where the data may be too sparse to produce reliable estimates based on the `local' subpopulation likelihood alone. If the data in a subpopulation are informative, then estimates of $\lambda_{i,q}$ will be based to a large extent on the data in that particular subpopulation. The less informative the data in a subpopulation, the more $\lambda_{i,q}$ will be influenced by information from `similar' subpopulations.
 
In related demographic work, parameters of hierarchical components such as (\ref{eq:weights}) are often implicitly approximated using a two-step procedure, where estimates of $\Phi_q$ and $\lambda_{i,q}$ are extracted using an SVD or PCA in a first step and then used as input to time series or regression models in a second step. In this paper, a joint estimation approach is adopted instead. Importantly, this enables leveraging both the `local' subpopulation information and the information obtained from `similar' subpopulations during model estimation. Combining both sources of information stabilizes estimates in noisy subpopulations. In addition, the joint hierarchical estimation approach allows for valid uncertainty quantification. 
 
The priors on $\bm{\delta}$ and $\bm{\beta}_q$ are specified as zero-centred horseshoe priors (\citealp{car-etal:hor}). This informative prior has most of its mass at zero and only allows the coefficients in $\bm{\delta}$ and $\bm{\beta}_q$ to deviate from zero if the data are sufficiently informative. This provides a variable selection and regularization mechanism, effectively reducing concerns about overfitting in contexts with noisy data. Regularizing $\bm{\beta}$ and $\bm{\delta}$ towards zero implies that the level of the subpopulation-specific process (governed by $\bm{\delta}$) and the shape of the subpopulation-specific process (governed by $\bm{\beta}_q$) are only affected by covariates where the data are clearly informative about these effects. To complete the prior setup, inverse gamma priors are chosen on the remaining variance parameters, with shape and rate parameters set to small constants. 

\subsection{Summary}

In summary, the main mechanisms of the proposed model can be conceptualized as follows. The $Q$ latent functions $\Phi_q(x)$ summarize information on age-specific patterns that are common to many (or all) subpopulations. The functions $\Phi_q(x)$ are further penalized to ensure smoothness in the age dimension. Given $\Phi_q(x)$, the conditional mean of $z_{i,x}$ is fully determined by the intercepts $\alpha_i$, the loadings $\lambda_{i,q}$ and the exposure term $O_{i,x}$. For $\lambda_{i,q}$ and $\alpha_i$, a hierarchical regression structure is used to share information between `similar' subpopulations, using the observed covariates $\bm{w}_i$ to measure similarity. The modelling approach therefore reflects all three key assumptions discussed in Section~\ref{sec:literature}.

The estimate of the demographic process in a subpopulation is then based on combining the information about $\alpha_i$ and $\lambda_{i,q}$ contained in the `local' subpopulation likelihood with information on those parameters obtained `globally' from `similar' subpopulations. The informativeness of the data determines the extent to which the `local' subpopulation information influences the final estimates relative to the `global' information from other subpopulations. In subpopulations with informative data, the effect of information-sharing may be overwritten, whereas, in subpopulations with noisy data, the information-sharing effect will be more pronounced. A discussion of model identification and parameter estimation using Markov chain Monte Carlo (MCMC) methods is given in Section~\ref{sec:computation}.

\section{Case Study: International Migration to Austria}
\label{sec:application}

\subsection{Data}

Register data on international immigration flows to Austria are analysed to illustrate the proposed modelling framework. The data are obtained from the Austrian national statistical office (\textit{Statistik Austria}) and consist of counts of incoming international migrants, summed over the period 2016-2020, and reported by individual years of age ($x = 0, \dots, 95$), sex and country of origin of the immigrants. Immigrant status is defined as being officially registered in Austria for more than 90 days. The data encompass all registered international immigrants and are derived from the Austrian central register of residents. It is available through custom extracts from \textit{STATCube}, the online data repository of \textit{Statistik Austria}. After excluding subpopulations with fewer than 10 observed immigrants, the data include immigrants from 155 countries of origin and a total of 300 subpopulations. 

As covariates $\bm{w}_i$, 15 binary indicators for the region of origin are included. These indicators are based on the UN sub-regional classification of countries. Additional indicator variables capture contiguity to Austria and EU membership of the country of origin, as well as the sex of migrants in a given subpopulation. Moreover, gross national income per capita, the number of battle deaths, population size, and working-age population share are included, each with respect to the origin country. Finally, the covariates include the number of migrants from a given country residing in Austria as of 2016, and the distance from that country's capital city to Vienna. These variables enter the model after a logarithmic transformation, except for the share of the working-age population. All continuous variables enter the model with an additional quadratic term to allow for non-linear effects. This set of observables is in the spirit of classical gravity models of migration (\citealp{beine2016practitioners}). The covariates cover many important theoretical channels driving international migration, such as economic opportunity, social networks, and fleeing violence (\citealp{czaika2022migration}). In addition, this set of variables allows for a rich information-sharing structure that facilitates borrowing information across different subpopulations based on regional similarities as well as similarities in living standards, conditions in the country of origin, and network ties to Austria. Summary statistics are provided in Tab.~\ref{table:summarystats} in the supplementary material.

\subsection{Computation and Model Selection}

Posterior inference is based on 50,000 posterior samples stored after an initial burn-in period of 25,000 iterations. Every 5th draw is saved to thin the posterior chain. This takes around 20-25 minutes on a single core of an AMD Ryzen 5 5500U. MCMC convergence behaviour is in general satisfactory, but requires some attention as data augmentation in a Poisson log-normal framework with many zero counts leads to a certain amount of posterior autocorrelation; see the supplementary material for further discussion and illustration (Section~\ref{sec:computation}; Fig.~\ref{fig:convergence}). 

Given the absence of a suitable offset term, we set $O_{i,x}=0$ for all $i$ and $x$. Cubic splines are used as base functions, with knots placed every five years from age 5 to 65, and boundary knots at ages 0 and 95. This reflects less dynamics in migration intensity at later ages and results in a reasonable balance between model flexibility and smoothness of the estimated demographic functions. A future avenue of research is to investigate or estimate an `optimal' placement of knots in different demographic modelling contexts.

To determine a suitable number of latent functions $Q$, model runs varying Q in $\{1, \dots, 10\}$ were considered in an exhaustive leave-one-curve-out cross-validation exercise and several predictive scores were compared. Details are provided in Section~\ref{sec:crossvalidation}.  Based on these results, we proceed to discuss results using $Q=6$, which provides the best accuracy-parsimony trade-off across all considered criteria. This is slightly higher than the number of factors in related papers on mortality modelling, such as \citet{clark2019general} and \citet{alexander2017flexible}, suggesting a higher degree of heterogeneity in the considered migration data. Results are largely similar between $Q=5$ and $Q=10$. When setting $Q<5$, the model loses the flexibility to adequately represent highly heterogeneous subpopulations, see Fig.~\ref{fig:varying_q} in the supplementary material for examples.

\subsection{Illustration of Model Fit}

The $Q=6$ extracted latent functions are shown in Fig.~\ref{fig:factors}.
The first factor captures patterns similar to a Rogers-Castro model migration schedule (\citealp{rogers1978model}) while the remaining factors mainly modulate migration intensity for certain age groups. The corresponding posterior means of the loadings $\lambda_{i,q}$, shown in Fig.~\ref{fig:loadings}, 
, indicate that all six factors are relevant for informing the shape of the age curves in many subpopulations. The first factor stands out as the most significant, reflected in the significant absolute size of the loadings.

\begin{figure}
     \centering
     \includegraphics[width=\textwidth]{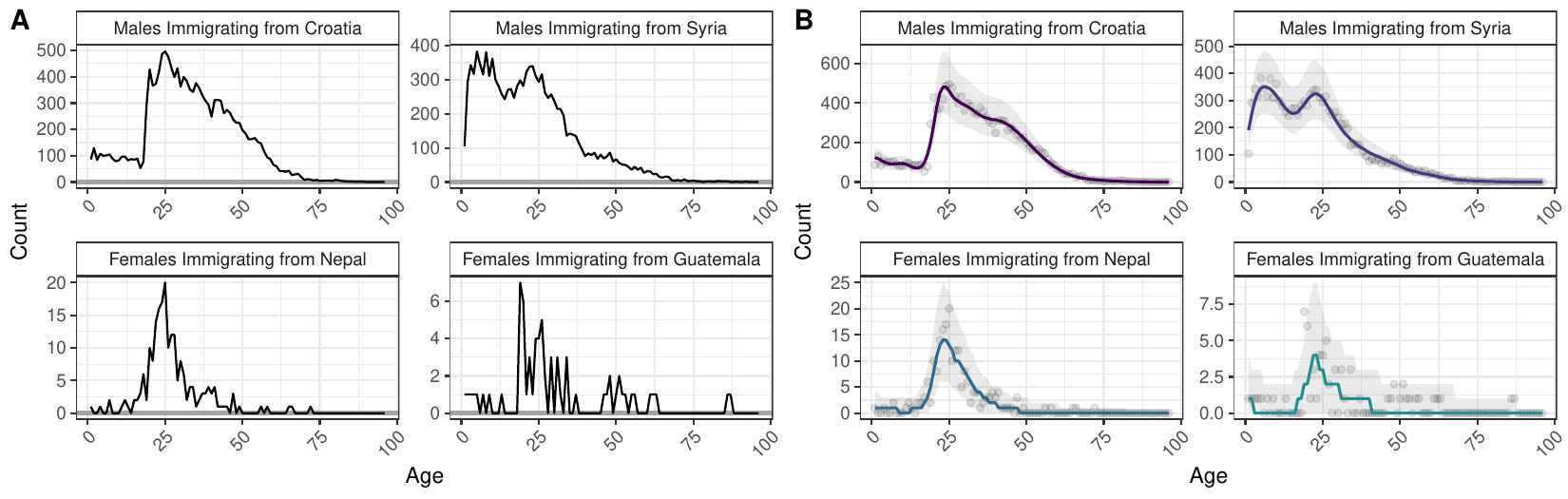}
       \caption{Raw (A) and estimated (B) age-specific counts. Shaded areas correspond to 95\% credible intervals. Points correspond to observed data. Upper left: Males migrating from Croatia. Upper right: Males immigrating from Syria. Bottom left: Females immigrating from Nepal. Bottom right: Females immigrating from Guatemala. Note that the scales of the $y$-axes differ between panels.}
        \label{fig:raw_vs_est}
\end{figure}

Fig.~\ref{fig:raw_vs_est} provides a visual representation of the raw data and model-based estimates of age-specific counts in four example subpopulations. These subpopulations include male migrants from Croatia (14,359 observed migrants in total), male migrants from Syria (11,642), female migrants from Nepal (204), and female migrants from Guatemala (68). Panel (A) shows the raw data, illustrating some of the heterogeneity, sparsity, and stochastic variation present in the age-specific immigration counts. 


In panel (B), the raw counts are overlaid with the estimation results, including probabilistic uncertainty bounds. The estimated mean of the predictive distribution of $y_{i,x}$ is shown as a smoothed model fit. In many applications, such model-based smooth estimates of the underlying demographic process are already the final output of interest. For instance, model-based estimates allow tracking of demographic patterns and trends in many subpopulations jointly or can serve as inputs to demographic projection models. By visual inspection of the results in Fig.~\ref{fig:raw_vs_est}, the model appears flexible enough to capture heterogeneity while remaining robust to the high level of noise in the input data. This will be confirmed more systematically using simulation studies in Section~\ref{sec:simulations}.  

\subsection{Testing for Heterogeneity Between Subpopulations}

In many cases, researchers are interested in exploring and making formal statements about homogeneity or heterogeneity of subpopulations. Such an exercise may serve as a purely exploratory tool or may be motivated by a desire to test hypotheses derived from theoretical models. The results may, in turn, inform projection models and initiate the development of new theoretical insights. 

\begin{figure}
     \centering
    
         \includegraphics[width=\textwidth]{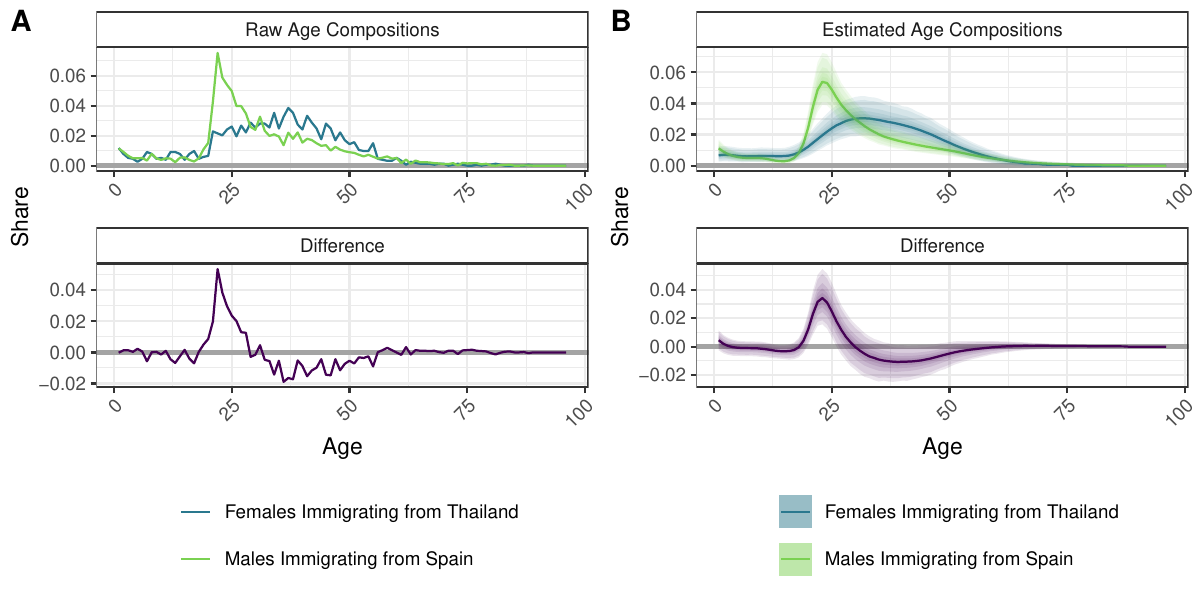}
        \caption{Raw (A) and estimated (B) age compositions of females immigrating from Thailand (blue), males immigrating from Spain (green) and the difference between the two age compositions (purple). Shaded areas correspond to 95\% Bayesian credible intervals.}
        \label{fig:diffs}
\end{figure}

Consider the problem of estimating whether there is a systematic difference in the age composition of two migrant subpopulations.\footnote{The age \textit{composition} (as opposed to the raw counts) is a natural object of interest when comparing the characteristics of migrants between subpopulations in settings where the respective levels of the counts are different. In addition, age compositions are key quantities in some demographic frameworks, e.g., for ex-post distribution of known migrant count aggregates in the absence of observed age-specific data.} More formally, suppose the goal is to make inferential statements about $(y_{j,x} / \sum_x y_{j,x}) - (y_{k,x} / \sum_x y_{k,x})$ for some $j$ and $k$ with $j \neq k$. To illustrate, consider Fig.~\ref{fig:diffs}, which compares the age composition of male Spanish immigrants and female Thai immigrants. The age compositions computed from the raw counts $y_{i,x}$ (top) as well as the difference between the two subpopulations (bottom) are shown in panel (A). It appears that the proportion of male Spanish migrants in their early 20s is larger than that of female Thai migrants. However, it is difficult to draw statistically valid conclusions from the raw age compositions alone. This is because it is inherently unclear whether the visually observed differences are due to systematic variation or due to noise. The Bayesian approach is highly useful in this context, as uncertainty bounds for functions of $y_{i,x}$ (such as age compositions) can be easily obtained via Monte Carlo simulation.


Panel (B) shows the smoothed estimated age compositions (top) and the difference between the smoothed curves (bottom), including 95\% credible intervals. From this analysis, it becomes clear that the difference observed in the raw data is likely to be a systematic phenomenon rather than an artefact of stochastic variation. In addition, there appears to be a slightly higher proportion of migrants aged 35-50 in the Thai female subpopulation. However, this difference is not estimated with fully conclusive precision, as indicated by the width of the uncertainty bands.

\subsection{Exploring Drivers of Heterogeneity: Linear and Nonlinear Covariate Effects}
\label{sec:covariates}

In addition to testing for the presence of heterogeneity, there is usually considerable interest in exploring potential drivers of heterogeneity. Equation (\ref{eq:weights}) implies that the model can be used to make statements about the partial effects of specific covariates in $\bm{w}_i$ on the shape and level of outcomes via their effects on $\lambda_{i,q}$ and $\alpha_i$. Consequently, the model has the capability to test theories on demographic behaviour and provide insights into the impact of policies and contextual variables on demographic outcomes.

Since the functions $\Phi_q(x)$ are constrained to have a mean of zero, the level effects are entirely determined by $\bm{\delta}$. Estimates of these level effects are given in Fig.~\ref{fig:level_fx} 
in the supplementary material. The indicator for EU membership of the sending country and the linear term of the pre-existing migrant population from a sending country have the clearest positive effects on the level of the immigration process. These reflect the free movement of EU citizens within the EU and the importance of network effects in international migration.

To assess the partial effect of covariates on the \textit{shape} of the immigration process, the model can be rewritten as a random effects function-on-scalar regression model, see \citet{kowal2020bayesian}. The partial shape effect of a covariate $w_{i,j}$ on $z_{i,x}$ is then given by the quantity $\sum_q \Phi_q(x) \beta_{q,j}$, where $\beta_{q,j}$ corresponds to the coefficient of the covariate $j$ in the $q$-th regression equation of the loadings $\lambda_{i,q}$. Again, the Bayesian approach allows to easily construct uncertainty bounds for this quantity based on Monte Carlo simulation. 

\begin{figure}
     \centering
         \includegraphics[width=\textwidth]{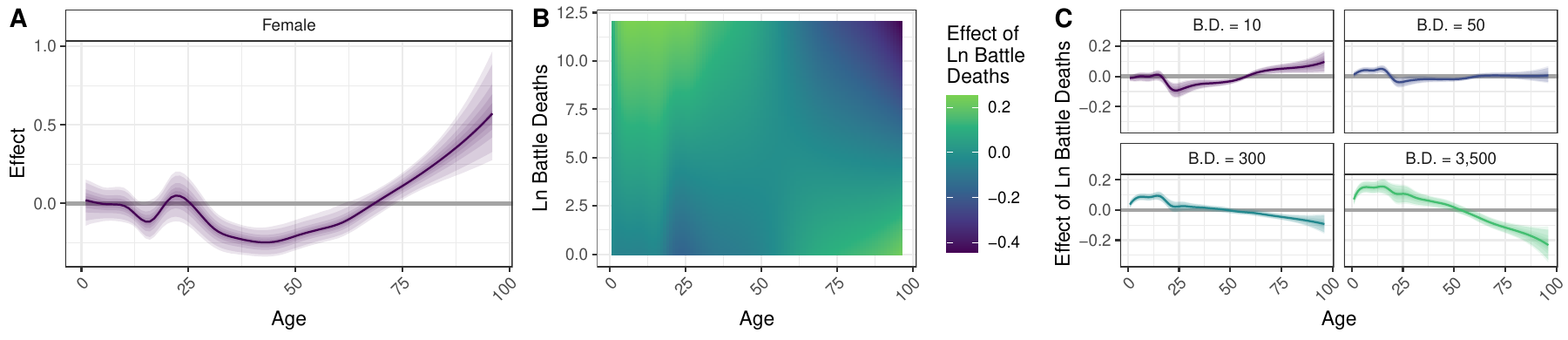}
        \caption{Examples of linear and non-linear covariate effects on the shape of the immigration process. Panel (A) shows the shape effect of the binary indicator for female subpopulations. Panel (B) shows a non-linear effect surface with point estimates of the impact of battle deaths in the country of origin on immigration flows. Panel (C) shows the impact of battle deaths in the country of origin on immigration flows for selected values of battle deaths. The shaded areas correspond to 95\% credible intervals.}
        \label{fig:effects_examples}
\end{figure}

Examples of such shape effect estimates are given in Fig.~\ref{fig:effects_examples}. Panel (A) shows the posterior distribution of the shape effect of the binary indicator for female subpopulations. These results imply that, on average and holding other factors constant, there are fewer female immigrants between the ages of 30 and 60 and more female immigrants at older ages relative to males. A plot of all estimated covariate effects on the shape of the immigration process can be found in Fig.~\ref{fig:shape_fx} 
in the supplementary material. The regularizing effect of the horseshoe prior on $\bm{\beta}$ is visible, as the shape effect of some covariates - e.g., the regional intercepts for Australia/New Zealand and Northern America - are estimated to be very close to zero for all ages. This implies that, on average, age-specific immigration patterns from these regions are not substantially different from the respective baseline estimates. For completeness, Fig.~\ref{fig:beta_factors} 
in the supplementary material provides the posterior mean estimates of all coefficients $\bm{\beta_{q}}$.


For variables that enter the equation of $\lambda_{i,q}$ with a linear and a quadratic term, the partial effect of a covariate $w_{i,j}$ on $z_{i,x}$ is given by $\sum_q \Phi_q(x) \beta_{q,j} + 2 w_{i,j} \sum_q \Phi_q(x) \beta_{q,n}$ where $\beta_{q,j}$ is the coefficient of the linear term $w_{i,j}$ in the $q$-th equation and $\beta_{q,n}$ is the coefficient of the quadratic term $w_{i,j}^2$ in the $q$-th equation.\footnote{While the use of quadratic terms to model non-linear effects is a simplistic approach, extensions to more complex non-linear modelling techniques such as spline expansions of covariates can be analysed on similar grounds.} Such non-linear effect estimates can be conveniently summarized in an effect surface. Panel (B) in Fig.~\ref{fig:effects_examples} shows the effect of (log) battle deaths in the country of origin on the shape of the immigration curve (colour) as a function of age ($x$-axis) and the level of battle deaths ($y$-axis). Finally, panel (C) of Fig.~\ref{fig:effects_examples} shows selected shape effects, evaluated at 10, 50, 300, and 3,500 battle deaths, roughly the 20th, 40th, 60th, and 80th quantiles of the log battle death distribution. The overall pattern is that as battle deaths in the origin country increase, immigration flows from this country to Austria are modulated, on average, towards more younger and fewer older immigrants. This shape effect becomes stronger with increasing levels of battle deaths. The estimates therefore imply a pronounced increase in emigration rates of children and teenagers following conflict events. This reflects the empirical pattern of young immigrants moving to Austria as refugees from conflict-affected countries.

\subsection{Probabilistic Projections and Counterfactual Exercises}
\label{sec:scenario}
\begin{figure}
\centering
        \includegraphics[width=\textwidth]{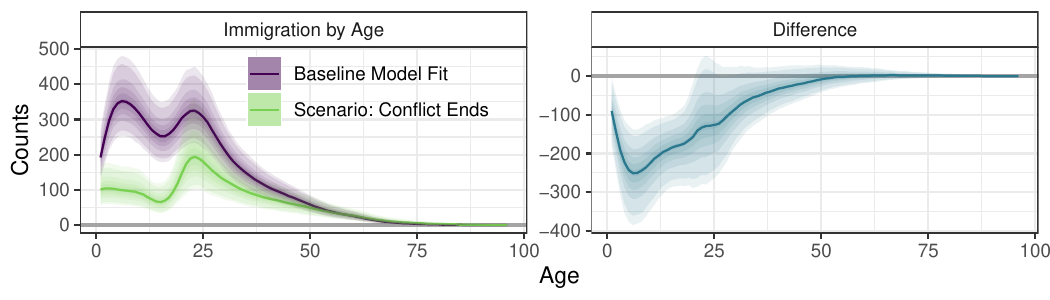}
        \caption{Scenario exercise. In-sample baseline fit for the Syrian male subpopulation (purple) versus model predictions for the Syrian male subpopulation with battle death indicators set to zero (green), as well as their difference (blue). Shaded areas correspond to 95\% credible intervals.}
        \label{fig:scenario}
\end{figure}

The ability to model the effect of covariates on both the level and shape of the immigration process also lends itself to probabilistic projections and counterfactual exercises, which are common tasks in empirical demography. Such exercises aim to explore likely demographic patterns conditional on given scenarios for changes in $\bm{w}_i$. A small empirical exercise is implemented to illustrate the capabilities of the model in this context. Suppose one is interested in estimating how immigration flows of male Syrians to Austria might be affected if violent conflict in Syria were to end. Model-based answers to this question can be explored by comparing the baseline fit of the model with a counterfactual prediction for a given subpopulation, where the observed covariate vector is swapped with a covariate vector corresponding to the scenario narrative.

The results of this exercise are given in Fig.~\ref{fig:scenario}. The figure shows in purple the baseline estimate for the subpopulation of males immigrating from Syria to Austria. In green, the figure shows the predictions of the model when the combat-related indicators in the covariate vector of the Syrian male subpopulation are set to zero.\footnote{It should be noted that this approach to a scenario capturing the end of a large-scale conflict is rather simplistic, and does not take into account potential relocation effects on migrants already residing in Austria, effects on per capita income, etc. If desired, this type of scenario exercise can be made arbitrarily complex, given suitable scenario paths for the covariates.} The difference between the two scenarios is shown in blue. The estimated difference and associated uncertainty bounds indicate that the number of migrants under the age of 20 is expected to decrease in the end-of-conflict scenario relative to the baseline. This reflects, by construction, the estimated relationship of conflict and age-specific migration; see Section~\ref{sec:covariates}.

\section{Predictive Ability and Comparison to Alternative Approaches}
\label{sec:simulations}

\begin{figure}
    \centering
    \includegraphics[width = \textwidth]{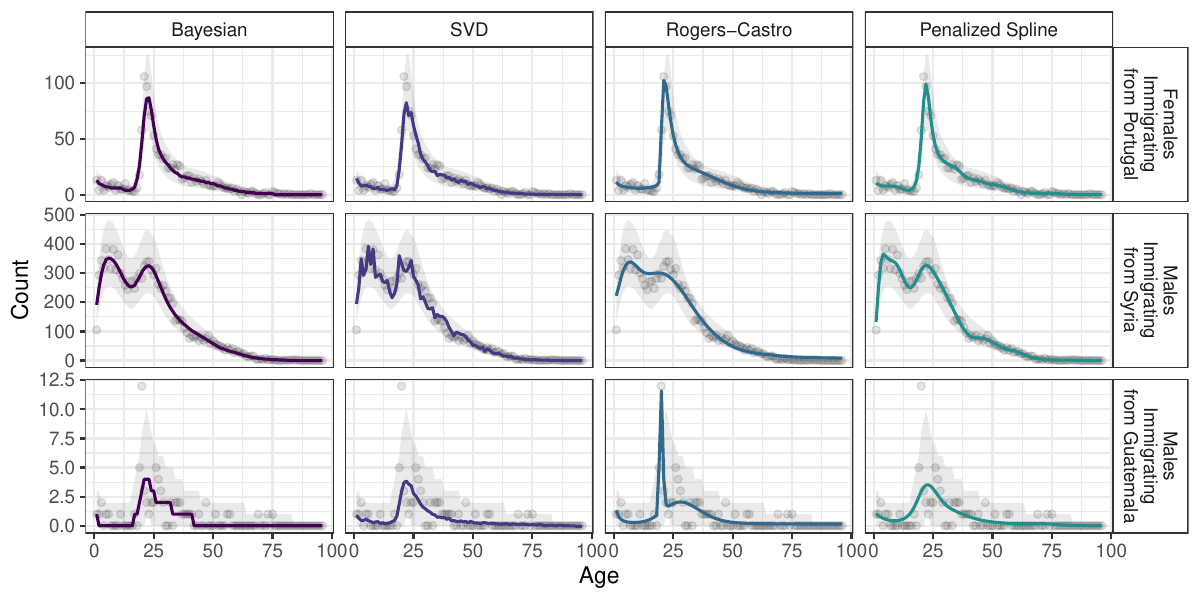}
    \caption{Selected in-sample fits obtained by four different modelling frameworks (columns) for three different subpopulations (rows). Shaded areas correspond to 95\% credible intervals from the Bayesian model. Points correspond to observed data. Refer to the text for details.}
    \label{fig:example_comparison}
\end{figure}

To assess the comparative performance of the proposed modelling framework and to highlight its predictive power, this section provides several illustrative comparisons with related modelling frameworks. First, selected in-sample estimates are discussed to outline some general insights. Second, further insights into the in-sample and out-of-sample predictive performance of different modelling frameworks are provided through systematic simulation studies. Finally, a cross-validation exercise is carried out using the Austrian immigration data.

\subsection{Insights on In-Sample Performance Using Selected Examples}

To provide some initial insights, three common alternative modelling frameworks are compared to the Bayesian model. For each model, the goal is to provide subpopulation-specific estimates of age-specific migration, based on the Austrian immigration data. First, an SVD-based reconstruction is computed, based on the first six singular vectors of the $\log(1+x)$ transformed counts. Second, results from a local smoother based on a Poisson penalized spline model, as in \citet{camarda2012mortalitysmooth}, are obtained. Third, the Rogers-Castro model (\citealp{rogers1978model}), a traditional demographic method for reconstructing migration schedules by age, is applied to the data.\footnote{The penalized spline model is fitted using the function \texttt{Mort1Dsmooth} in the defunct R package \texttt{MortalitySmooth}. In the absence of an available procedure to fit the Rogers-Castro model directly to count data, the Rogers-Castro model is fitted by first transforming the count data to an age composition, then fitting the compositional data with a Gaussian likelihood model using the R package \texttt{rcbayes}, and then transforming back to the count scale based on median modelled estimates of the age composition.} 

Results for three selected subpopulations are shown in Fig.~\ref{fig:example_comparison}. The top row shows the fit for a subpopulation with relatively large counts and a regular migration pattern, representing an informative setting. Roughly similar results are obtained from all four models. The second row shows the fit for a subpopulation with a large sample size but a relatively `irregular' pattern. Here, the Rogers-Castro model is not flexible enough to account for the heterogeneity of the pattern and underfits. This happens because the parametric function underlying the Rogers-Castro model is based on the shape of `regular' age-specific migration patterns. In comparison, the SVD model, spline smoothing, and the Bayesian model give similar results, with the SVD model capturing significantly more noise. The last row shows the fit for a subpopulation that is relatively regular in shape, but with a small sample size. The resulting fits show that the SVD and Bayesian approaches, as well as the spline model, can still recover a reasonable age pattern despite the extremely noisy data. In contrast, the Rogers-Castro model provides a sub-optimal fit in this setting, clearly overfitting the noisy data.

These preliminary investigations broadly illustrate some comparative advantages of the Bayesian approach relative to the alternative modelling frameworks. The Bayesian model combines the idea of underlying common components from the SVD and Rogers-Castro models, as well as the idea of smoothing from the penalized spline framework. This allows the Bayesian model to overcome the inflexibility of the Rogers Castro framework, while remaining more robust to noise than the SVD framework. Compared with local spline smoothing methods, the Bayesian approach allows for information sharing, further improving efficiency, robustness of the model fit and additionally providing a mechanism for counterfactual analysis, projections and predictions.

\subsection{Systematic Simulation Studies Based On Synthetic Data}

To corroborate the selected results from the previous subsection, a systematic simulation study is carried out to evaluate the performance of the proposed model. The purpose of this simulation study is to identify scenarios in which the additional complexity of the Bayesian model pays off relative to simpler modelling frameworks. Three different exercises are considered. First, the accuracy of \textit{in-sample} estimates is assessed, providing insight into the overall quality of the smoothed model fits (Fig.~\ref{fig:insample_baseline}). 
Second, prediction quality when \textit{imputing partially missing data} is assessed to explore the utility of the model in scenarios with incomplete data (Fig.~\ref{fig:missing_data}). 
Third, \textit{out-of-sample predictions} are evaluated in settings where demographic data are missing completely for some subpopulations (Fig.~\ref{fig:oos_evaluation}). 
In all exercises, the aim is to estimate a known systematic signal $\alpha_i+z_i(x)$ from noisy observations $y_{i,x}$. 

The synthetic data sets are simulated as follows. First, the five leading principal components are extracted using an SVD of the $\log(1+x)$ transformed Austrian immigration counts. These principal components are then smoothed using a B-spline basis expansion with 7 equally spaced interior knots. These represent the ground truth of the underlying functions $\Phi_q(x)$.\footnote{An additional scenario considers data that features abrupt jumps at ages 18-20 years in randomly selected 5\% of subpopulations to explore model behaviour in scenarios with non-smooth outliers (Fig.~\ref{fig:insample_outliers}).} 
Ten covariates and all elements of $\bm{\beta}_q$ are simulated from $\mathcal{N}(0,1)$. The intercept term in $\bm{\delta}$ is set to 15 and all other elements of $\bm{\delta}$ are simulated from $\mathcal{N}(0,1)$. Large positive counts $y_{i,x}$ result from setting the intercept in $\bm{\delta}$ to 15. This makes $z_{i,x}$ essentially equivalent to $\log(y_{i,x})$. This is done to minimize the effect of the frequently used $log(1+x)$ transformation of counts or rates before the estimation of SVD-based models. The results of these scenarios can therefore be seen as an upper bound on the performance of competing models that rely on such log transformations.\footnote{An additional scenario with $\delta = -1.5$ is considered to explore the impact of data sets with an abundance of zero counts (Fig.~\ref{fig:insample_sparse}).} 
Three random elements of $\bm{\delta}$ and three random elements of $\bm{\beta}$ are set to zero to simulate a scenario where some covariates are uninformative predictors. Finally, all $\sigma_{\lambda,q}=1$ and $\sigma^2_\alpha = 0.5$ are fixed. In most settings, we consider low noise ($\sigma^2 = 0.1$) and high noise ($\sigma^2 = 1$) scenarios. For the partially missing data exercise, one observation per age curve is randomly selected and dropped from the training data. Each simulation is repeated 25 times. Root mean square errors, mean absolute errors, and mean percentage errors of all competing models relative to the ground truth are recorded, averaged over the 25 replications.

As competing models for the in-sample simulation, we consider a least squares reconstruction based on the first five principal components of the log counts using SVD, a penalized spline smoothing procedure on the log counts independently applied to each subpopulation, and a procedure that first smooths the log counts using penalized splines and then computes a least squares reconstruction using the first five principal components extracted from the smoothed series. In addition, a variant of the TOPALS framework is estimated where deviations from a reference age pattern are modelled using penalized splines (\citealp{de2011new}; \citealp{dyrting2020smoothing}). As reference pattern, we use the average age pattern across all subpopulations. For penalized spline models, hyperparameters are selected to minimize the generalized cross-validation criterion using the function \texttt{smooth.Pspline} in the R package \texttt{pspline}. For imputing partially missing data, we consider linear interpolation of missing log counts, penalized spline interpolation of the missing log counts and the outlined TOPALS variant as competing frameworks.\footnote{The Rogers-Castro model and the Poisson count smoothing framework of \citet{camarda2012mortalitysmooth} are not explicitly considered, as both can be difficult to fit in sparse and noisy multipopulation environments, leading to repeated software failures, suboptimal fits, and uncompetitive predictive performance.} Finally, for the out-of-sample prediction exercise, we consider as competitors two models that learn linear regression functions in the loadings $\lambda_{i,q}$ and intercepts $\alpha_i$. The intercepts $\alpha_i$ are set to the average log count of subpopulation $i$, and $\lambda_{i,q}$ are estimated from an SVD of the demeaned log counts, once with and once without prior smoothing of the log counts. 

The results of these simulation exercises can be summarized as follows. First, the Bayesian model improves over all competing models in all considered settings (Fig.~\ref{fig:insample_baseline}-\ref{fig:insample_sparse}). 
This implies that when the three assumptions in Section~\ref{sec:literature} are assumed to hold, incorporating them jointly leads to clear performance gains relative to simpler methods that leverage these assumptions only partially. Second, the performance of the various methods is not very different when noise is low, but diverges when noise is high (see Fig.~\ref{fig:insample_examples_sim} 
for an example) and when sparsity in the count outcomes increases. This suggests that local smoothers and SVD-based methods are useful modelling tools in settings where subpopulations are characterized by large counts and where noise is not a severe issue. However, in the presence of subpopulations with small sample sizes and noise, the Bayesian approach significantly improves over alternative methods. Third, when imputing missing data, the Bayesian model outperforms basic interpolation rules already in the simplest settings, where noise is low and only single data points are missing. These gains are likely to be even more pronounced in noisy data or when larger sets of data points are missing. 

\subsection{Cross-Validation Exercise Based On Austrian Immigration Data}
\label{sec:crossvalidation}

In addition to the systematic exercises using simulated data, a real data out-of-sample prediction exercise is carried out using the Austrian immigration data. The double purpose of this is to gain further insight into the comparative predictive power of the framework and to aid the selection of an appropriate value of $Q$ for the motivating data set. A leave-one-curve-out cross-validation approach is implemented where, in each run, the 300 observed subpopulations are divided into 299 training subpopulations and a single hold-out subpopulations on which the model predictions are evaluated. Each subpopulation was therefore part of a hold-out sample after completing all runs. The Bayesian model is estimated with $Q$ varying in $\{1, \dots, 10\}$ in each of the five runs. In addition, we consider estimating the Bayesian model without any covariates present to evaluate the importance and relevance of the covariates in describing the data at hand. Predictions from two competing two-stage approximation approaches are also collected. The first stage is an SVD on the demeaned log-transformed counts, once with and once without prior smoothing of the log counts. In the second stage, linear regression models with SVD-based approximations of $\lambda_{i,q}$ and $\alpha_i$ as outcomes are estimated using ordinary least squares. Model predictions are then obtained by predicting $\lambda_{i,q}$ and $\alpha_i$ out-of-sample and combining these predictions with the appropriately scaled singular vectors. All models assume $\log(1+y_{i,x})$ as the outcome of interest for numerical stability and in favour of SVD-based methods that rely on a transformed count outcome. 

A summary of the results is provided in Tab.~\ref{table:crossvalidation}, which shows the out-of-sample root mean squared errors, mean absolute errors, and correlations between predicted and true values. These metrics are averaged across 300 predictive runs. We find that the SVD approaches, both with and without prior smoothing of the counts, perform rather similarly. The Bayesian model outperforms the SVD approaches for all values of $Q$ considered. Inclusion of the covariates $\bm{w}_i$ in the modelling framework significantly improves out-of-sample predictive power of the framework. The performance of the Bayesian model with covariates improves considerably with the first three factors, after which the marginal effect of an additional factor becomes smaller. Raising the rank beyond $Q=6$ produces no systematic improvement in any cross-validation metric; this specification is examined in detail in Section~\ref{sec:application}.

\begin{table}[t]
\caption{Results of Leave-One-Curve-Out Cross-Validation Exercise.}
\begin{center}
\resizebox{\textwidth}{!}{%
\begin{threeparttable}
\begin{tabular}{llrrrrrrrrrr}
\toprule
 &  & Q = 1 & Q = 2 & Q = 3 & Q = 4 & Q = 5 & Q = 6 & Q = 7 & Q = 8 & Q = 9 & Q = 10\\
\midrule
RMSE & Bayesian & 0.501 & 0.492 & 0.484 & 0.483 & 0.480 & 0.479 & 0.479 & 0.478 & 0.479 & 0.478\\
 & SVD & 0.535 & 0.516 & 0.508 & 0.506 & 0.505 & 0.503 & 0.503 & 0.503 & 0.504 & 0.504\\
 & Smooth SVD & 0.536 & 0.516 & 0.508 & 0.506 & 0.505 & 0.503 & 0.503 & 0.503 & 0.503 & 0.503\\
 & Bayesian (No Covariates) & 1.215 & 1.217 & 1.214 & 1.214 & 1.214 & 1.210 & 1.211 & 1.209 & 1.210 & 1.209\\
\addlinespace
MAE & Bayesian & 0.379 & 0.372 & 0.365 & 0.363 & 0.363 & 0.361 & 0.361 & 0.361 & 0.361 & 0.361\\
 & SVD & 0.422 & 0.397 & 0.393 & 0.391 & 0.390 & 0.387 & 0.388 & 0.388 & 0.388 & 0.388\\
 & Smooth SVD & 0.422 & 0.397 & 0.394 & 0.390 & 0.390 & 0.387 & 0.387 & 0.387 & 0.387 & 0.387\\
 & Bayesian (No Covariates) & 1.044 & 1.045 & 1.043 & 1.043 & 1.043 & 1.040 & 1.041 & 1.039 & 1.040 & 1.039\\
\addlinespace
Corr. & Bayesian & 0.773 & 0.776 & 0.780 & 0.780 & 0.782 & 0.782 & 0.782 & 0.782 & 0.782 & 0.783\\
 & SVD & 0.724 & 0.744 & 0.749 & 0.751 & 0.751 & 0.752 & 0.752 & 0.752 & 0.752 & 0.751\\
 & Smooth SVD & 0.723 & 0.742 & 0.748 & 0.749 & 0.749 & 0.750 & 0.750 & 0.750 & 0.750 & 0.750\\
 & Bayesian (No Covariates) & 0.756 & 0.755 & 0.756 & 0.755 & 0.756 & 0.756 & 0.756 & 0.756 & 0.756 & 0.756\\
\bottomrule
\end{tabular}
\begin{tablenotes}
\item \noindent \textit{Note:} Root mean squared error (RMSE), mean absolute error (MAE), and correlation (Corr.) between the posterior mean predictions and the true values from the cross-validation (CV) exercise are averaged over 300 curves. These metrics are calculated using log-transformed counts to ensure numerical stability.
\end{tablenotes}
\end{threeparttable}%
}
\end{center}
\label{table:crossvalidation}
\end{table}

\section{Concluding Remarks}
\label{sec:conclusion}

This paper considers the problem of joint modelling of age-specific counts in many, potentially small, demographic subpopulations. A Bayesian model is presented that combines three common assumptions from empirical demographic literature in a unified probabilistic framework. The model is based on the idea that latent and smooth underlying common age patterns can be extracted from the data and that information on the demographic process is shared between similar subpopulations. 

The method is applied to Austrian register data on age-specific immigrant flows in 300 subpopulations, based on sex and country of origin. This case study is used to demonstrate the main applications of the model, including obtaining smoothed estimates of the demographic process, making probabilistic statements about the presence of heterogeneity, exploring systematic drivers of heterogeneity, and probabilistic counterfactual exercises. Further illustrative and systematic predictive comparisons are used to benchmark the modelling framework against various competing models, demonstrating the generally good performance of the proposed model. Importantly, the model is flexible enough to describe systematic patterns of heterogeneity while remaining robust to noise through regularization mechanisms. 

This study confirms the considerable effectiveness of dimensionality reduction techniques in the age dimension of multipopulation demographic data. It is worth emphasizing that of the three model assumptions described, the assumption of latent underlying common patterns across multiple populations appears to be the most important for providing reasonable model estimates and predictions, based on the case study and systematic explorations of model performance. A relevant finding for practitioners is that SVD-based modelling is expected to work well in settings with large counts and relatively little noise (e.g. age-specific mortality at the country level). In such settings, SVD-based models produce reasonably accurate estimates and require less computation time than MCMC-based posterior approximation algorithms. Combined with a hierarchical structure exploiting temporal, spatial, or covariate-based proximity of subpopulations, powerful demographic modelling tools can be obtained. The assumption of smoothness across ages becomes increasingly important when data are noisy. `Local' smoothing methods without any information sharing mechanisms should be used with caution when dealing with many subpopulations, as there is a risk of inefficiency and overfitting. 

In terms of applied work, several avenues of future research might be interesting to explore. First, although we apply the model in a migration setting, applications to fertility and mortality data and comparisons with other standard modelling frameworks would be informative. Given that mortality and fertility schedules tend to be somewhat more regular than the international migration schedules analysed in this article, the modelling framework is expected to perform similarly or better in these contexts. Second, applications to partially incomplete data, as in the context of life tables in developing countries, appear promising. Third, it would be interesting to consider alternative dimensions for smoothing demographic outcomes. For example, the model can be extended to consider smoothing in the dimension of a continuous variable such as population density, to explore demographic phenomena along an urban-rural continuum. 

In terms of prior and modelling choices, the most impactful decisions are typically the choice of latent factors $Q$, the number and placement of spline knots $K$, and the assumption of second-order random walk dynamics for $f_{q,k}$. Regarding the first two, future research could focus on incorporating these decisions explicitly into the estimation process, as the uncertainty surrounding these choices can often be framed as a parameter estimation problem in the Bayesian paradigm. For instance, \citet{fruhwirth2024sparse} estimate $Q$ endogenously as a parameter during MCMC and \citet{dimatteo2001bayesian} discuss Bayesian spline regression when the number of knots and their placement are free parameters to be estimated. The second-order random walk assumption on the spline coefficients provides robustness to noisy patterns in the data, but can lead to oversmoothing of abrupt changes in demographic schedules, such as sudden spikes in migration schedules around age 18 when young adults migrate for education. Here, future research could consider combining smooth and non-smooth factors to capture both smooth and more abrupt changes in demographic schedules.

In addition, several methodological extensions are potentially worth investigating. First, extensions of the hierarchical part of the model to allow for smooth temporal components or spatial random effects are promising generalizations. In this vein, \citet{zens2025bayesian} considers a parsimonious Bayesian matrix factor framework for demographic analysis that jointly models latent age and time factors. Second, a more complex nonlinear regression approach for the priors on $\alpha_i$ and $\lambda_{i,q}$ (possibly based on another hierarchical layer including P-splines or regression trees) could further improve predictive performance. Finally, only a moderate number of subpopulations (i.e. a few hundred) are considered in the case study in this paper. Some interesting settings will potentially have thousands of very sparse subpopulations, for example when considering US county level data (\citealp{yu2023probabilistic}; \citealp{dharamshi2025jointly}). It would be informative and potentially lead to further model refinement and extensions to explore the performance and limitations of the model in such high-dimensional settings. 




\singlespacing
\vspace*{2em}\noindent{\large\MakeUppercase{\textbf{References}}}\vspace*{0.5em}
\printbibliography[heading=none]
\newpage

%

\setcounter{equation}{0}
\setcounter{figure}{0}
\setcounter{section}{0}
\setcounter{table}{0}
\renewcommand\thesection{S\arabic{section}}
\renewcommand\thefigure{S\arabic{figure}}
\renewcommand\thetable{S\arabic{table}}
\newrefsection

\vspace*{2em}\noindent{\large\MakeUppercase{\textbf{Supplementary Material}}}\vspace*{-0.5em}
\section{Model Identification and Bayesian Estimation Using MCMC}

\label{sec:computation}
\onehalfspacing

Equation (\ref{eq:z_factors}) implies that the model is only identified up to rotating, scaling, relabeling and changing the sign of the latent functions $\Phi_q(x)$. To partially address these identification issues and numerically stabilize the MCMC algorithm, the constraints from \citet{kowal2020bayesian} are implemented. This effectively fixes the system to an orthonormal rotation by conditioning the estimation process on $\Phi_j(x)'\Phi_j(x) = 1$ and $\Phi_j(x)'\Phi_k(x) = 0$ for $j \neq k$. This orthonormality constraint renders the model similar to a probabilistic principal components model under a Poisson likelihood, compare for example \citet{chiquet2018variational}. During posterior simulation, the signs and ordering of the latent functions are left unidentified, which was not an issue for MCMC sampling in the applications considered in this article. If sign switching becomes an issue, however, simple and efficient ex-post reordering schemes are available (\citealp{kastner2017efficient}). An order constraint on $\sigma^2_{\lambda,q}$ could be used to resolve the labeling invariance problem. 

Given the model specification and the identification scheme, the goal is then to obtain samples from the posterior distributions of the unknown parameters $\alpha_i$, $\bm{\delta}$, $\lambda_{i,q}$, $\bm{\beta}_q$, $\sigma^2$, $\sigma^2_\alpha$ and $\sigma^2_{\lambda,q}$, $\tau_q$, $\kappa_{q,k}$, as well as the spline coefficients $f_{q,k}$. The proposed estimation scheme is motivated by the fact that conditional on $z_{i,x}$, the remaining model is essentially a Gaussian regression model, making data augmentation (\citealp{tan-won:cal}) a natural candidate for posterior simulation. The specific algorithm after conditioning on $z_{i,x}$ follows along the lines of the algorithm described in \citet{kowal2020bayesian}. Lemma 1 in \citet{kowal2020bayesian} simplifies several sampling steps and relies on projected observations, which will be denoted as 

\[
z^*_{i,q} = \Phi_q(x)' \Bigl(\bm{z}_i - \alpha_i \bm{\iota}\Bigr), \quad \text{with } \bm{z}_i = \bigl(z_{i,1},\dots,z_{i,A}\bigr)'
\]

where $\bm{\iota}$ is a $A \times 1$ vector of ones. Posterior samples are obtained by setting appropriate initial values for all parameters and then iterating through the following steps.\footnote{For simplicity, offset terms are assumed fixed at $O_{i,x}=0$ in these posterior updates. Incorporating non-zero offsets would require only modest, straightforward modifications to the given sampling scheme.}

\bigskip

\noindent\textbf{Update latent outcomes \(z_{i,x}\).} For \(i=1,\dots,N\) and \(x=1,\dots,A\), update
\[
z_{i,x} \sim p(z_{i,x} \mid y_{i,x}, \alpha_i, \sigma^2, \Phi_q(x), \lambda_{i,q})
\]
with
\[
p\Bigl(z_{i,x}\mid \cdot\Bigr) \propto \mathcal{P}\Bigl(y_{i,x};\, e^{z_{i,x}}\Bigr)
\times \mathcal{N}\Bigl(z_{i,x};\, \alpha_i + \sum_{q=1}^Q \Phi_q(x)\,\lambda_{i,q},\, \sigma^2\Bigr),
\]
using an appropriate posterior sampling algorithm. We use univariate adaptive Metropolis updates in the style of \citet{roberts2009examples}. Gradient-based updates have also been used successfully (e.g., \citealp{steel2024model}) but typically come at a relatively high computational cost when many latent variables are involved. In case exact updating of \(z_{i,x}\) is too computationally expensive, several approximations to Poisson and Poisson lognormal models (e.g., \citealp{el1973bayesian}; \citealp{chan2009bayesian}) are available. For large counts, a simple approximation is to fix \(z_{i,x} = \log(y_{i,x})\) since, as \(y_{i,x}\to\infty\), we have \(z_{i,x} \to \log(y_{i,x})\) with probability one (see also \citealp{pavone2024learning}).

\medskip
\noindent\textbf{Update overdispersion parameter \(\sigma^2\).} The conditional posterior density
\[
p(\sigma^2 \mid z_{i,x}, \alpha_i, \Phi_q(x), \lambda_{i,q})
\]
is proportional to an inverse Gamma density,
\[
\sigma^2 \sim \mathcal{IG}(c_N, C_N),\quad c_N = c_0 + 0.5\,NA,\quad C_N = C_0 + 0.5\sum_{i,x} \Bigl(z_{i,x} - \alpha_i - \sum_{q} \Phi_q(x)\,\lambda_{i,q}\Bigr)^2,
\]

where the prior hyperparameters $c_0$ and $C_0$ can be set to small constants.

\medskip
\noindent\textbf{Update intercept terms \(\alpha_i\).} For \(i=1,\dots,N\),
\[
p(\alpha_i \mid z_{i,x}, \Phi_q(x), \lambda_{i,q}, \sigma^2, \sigma^2_\alpha)
\]
is proportional to
\[
\alpha_i \sim \mathcal{N}(a_{N,i}, A_{N}),\quad A_{N}=\Bigl(\frac{1}{\sigma^2_\alpha}+\frac{A}{\sigma^2}\Bigr)^{-1},\quad a_{N,i}=A_{N}\Bigl(\frac{a_{0,i}}{\sigma^2_\alpha}+\frac{1}{\sigma^2}\sum_x\Bigl(z_{i,x}-\sum_q \Phi_q(x)\,\lambda_{i,q}\Bigr)\Bigr),
\]
with \(a_{0,i}=\bm{w}_i'\bm{\delta}\).

\medskip
\noindent\textbf{Update spline coefficients \(f_{q,k}\).} For each \(q=1,\dots,Q\), update \(\bm{f}_q=(f_{q,1},\dots,f_{q,K})'\) as follows. Compute the penalty matrix
\[
\bm{\Omega}_q = \bm{D}'\bm{K}_q\bm{D},
\]
where \(\bm{K}_q=\operatorname{diag}(\kappa_{q,k})\) and \(\bm{D}\) is a \((K-2)\times K\) matrix with entries
\[
d_{i,j}=\begin{cases} 1,& i=j,\\ -2,& j=i+1,\\ 1,& j=i+2,\\ 0,& \text{otherwise.} \end{cases}
\]
Define the working outcomes
\[
\tilde{z}_{i,x}=z_{i,x}-\alpha_i-\sum_{q'\neq q}\Phi_{q'}\,\lambda_{i,q'}.
\]
Then, the conditional posterior distribution of $\bm{f}_q$ is proportional to a multivariate Gaussian density
\[
\bm{f}_q \sim \mathcal{N}(\bar{\bm{f}}_{N,q},\bar{\bm{F}}_{N,q}),
\]
with
\[
\bar{\bm{F}}_{N,q}=\Bigl(\tau_q^{-1}\bm{\Omega}_q+\frac{1}{\sigma^2}\Bigl(\sum_i\lambda_{i,q}^2\Bigr)\bm{B}'\bm{B}\Bigr)^{-1},\quad
\bar{\bm{f}}_{N,q}=\bar{\bm{F}}_{N,q}\Bigl(\frac{1}{\sigma^2}\bm{B}'\tilde{\bm{Z}}'\bm{\Lambda}_q\Bigr).
\]
Here, \(\bm{B}\) is an \(A\times K\) matrix with columns \(\bm{B}_k\), \(\tilde{\bm{Z}}\) is the \(N\times A\) matrix with entries \(\tilde{z}_{i,x}\), and \(\bm{\Lambda}_q=(\lambda_{1,q},\dots,\lambda_{N,q})'\). After sampling an initial draw from this density, this candidate draw is immediately rotated and scaled to satisfy the identification constraints using the algorithm described in \citet{kowal2020bayesian}.

\medskip
\noindent\textbf{Update loadings \(\lambda_{i,q}\).} Using the model implied by (\ref{eq:weights})
\[
\lambda_{i,q}=\bm{w}_i'\bm{\beta}_q+\gamma_{i,q},\quad \gamma_{i,q}\sim \mathcal{N}(0,\sigma^2_{\lambda,q}),
\]

and following \citet{kowal2020bayesian}, the update of $\lambda_{i,q}$ is accomplished via updating of $\gamma_{i,q}$ and $\bm{\beta}_q$ in a marginalized projected model, and then setting $\lambda_{i,q} = \bm{w}'_i\bm{\beta}_q + \gamma_{i,q}$. By Lemma~1 of \citet{kowal2020bayesian}, the conditional posterior density of $\gamma_{i,q}$ is proportional to a Gaussian density
\[
\gamma_{i,q}\sim \mathcal{N}(g_{N,i,q},G_{N,q}),\quad G_{N,q}=\Bigl(\frac{1}{\sigma^2_{\lambda,q}}+\frac{1}{\sigma^2}\Bigr)^{-1},\quad g_{N,i,q}=G_{N,q}\Bigl(\frac{z^*_{i,q}-\bm{w}_i'\bm{\beta}_q}{\sigma^2}\Bigr).
\]

\medskip
\noindent\textbf{Update global spline smoothing coefficients \(\tau_q\).} For \(q=1,\dots,Q\), the posterior distribution of $\tau_q$ is proportional to an inverse gamma density
\[
\tau_q\sim \mathcal{IG}(d_q,D_q),\quad d_q=d_0+0.5\,(K-2),\quad D_q=D_0+0.5\,\bm{f}_q'\bm{\Omega}_q\,\bm{f}_q,
\]
where the prior hyperparameters $d_0$ and $D_0$ can be set to small constants.

\medskip
\noindent\textbf{Update local spline smoothing coefficients \(\kappa_{q,k}\).} For $q=1, \dots, Q$, the conditional posterior distributions of the weights $\kappa_{q,k}$ are proportional to a gamma densities
\[
\kappa_{q,k}\sim \mathcal{G}\Bigl(1,\,0.5+\frac{u_{q,k}^2}{2\tau_q}\Bigr).
\]

\medskip
\noindent\textbf{Update hierarchical regression parameters \(\bm{\beta}_q\).} Using Lemma 1 of \citet{kowal2020bayesian}, for \(q=1,\dots,Q\), the conditional posterior density of $\bm{\beta}_q$ is proportional to a multivariate Gaussian density
\[
\bm{\beta}_q\sim \mathcal{N}(\bm{b}_{N,q},\bm{B}_{N,q}),
\]
with
\[
\bm{B}_{N,q}=\Bigl(\frac{\bm{W}'\bm{W}}{\sigma^2+\sigma^2_{\lambda,q}}+\bm{B}_{0,q}^{-1}\Bigr)^{-1},\quad
\bm{b}_{N,q}=\bm{B}_{N,q}\Bigl(\frac{\bm{W}'\bm{z}^*_q}{\sigma^2+\sigma^2_{\lambda,q}}\Bigr),
\]
where \(\bm{z}^*_q=(z^*_{1,q},\dots,z^*_{N,q})'\) and \(\bm{W}\) is the \(N\times R\) matrix with rows \(\bm{w}'_i\). The prior covariance is 
$\bm{B}_{0,q}=\operatorname{diag}(\xi_{\beta,q} \rho_{\beta,q,1},\dots,\xi_{\beta,q}\rho_{\beta,q,R})$ where $\xi_{\beta,q}$ and $\rho_{\beta,q,r}$ are the global and local shrinkage parameters of the horseshoe prior specified for the elements of $\bm{\beta}_q =(\beta_{q,1}, \dots, \beta_{q, R})'$.

\medskip
\noindent\textbf{Update hierarchical regression parameters \(\bm{\delta}\).} The conditional posterior density of $\bm{\delta}$ is proportional to a multivariate Gaussian density
\[
\bm{\delta}\sim \mathcal{N}(\bm{d}_N,\bm{D}_N),\quad \bm{D}_N=\Bigl(\frac{\bm{W}'\bm{W}}{\sigma^2_\alpha}+\bm{D}_0^{-1}\Bigr)^{-1},\quad \bm{d}_N=\bm{D}_N\Bigl(\frac{\bm{W}'\bm{\alpha}}{\sigma^2_\alpha}\Bigr),
\]
with \(\bm{D}_0=\operatorname{diag}(\xi_{\delta}\rho_{\delta,1},\dots,\xi_{\delta}\rho_{\delta,R})\) where $\xi_{\delta}$ and $\rho_{\delta,r}$ are the global and local shrinkage parameters of the horseshoe prior specified for the elements of $\bm{\delta} = (\delta_1, \dots, \delta_R)'$.

\medskip
\noindent\textbf{Update hierarchical variance parameter \(\sigma^2_\alpha\).} The conditional posterior density of $\sigma^2$ is proportional to an inverse gamma density
\[
\sigma^2_\alpha\sim \mathcal{IG}(s_N,S_N),\quad s_N=s_0+0.5\,N,\quad S_N=S_0+0.5\sum_i\Bigl(\alpha_i-\bm{w}_i'\bm{\delta}\Bigr)^2,
\]
where the prior hyperparameters $s_0$ and $S_0$ can be set to small constants.

\medskip
\noindent\textbf{Update hierarchical variance parameters \(\sigma^2_{\lambda,q}\).} For \(q=1,\dots,Q\), the conditional posterior density of $\sigma^2_{\lambda,q}$ is proportional to an inverse gamma density
\[
\sigma^2_{\lambda,q}\sim \mathcal{IG}(l_N,L_N),\quad l_N=l_0+0.5\,N,\quad L_N=L_0+0.5\sum_i\Bigl(\lambda_{i,q}-\bm{w}_i'\bm{\beta}_q\Bigr)^2,
\]
where the prior hyperparameters $l_0$ and $L_0$ can be set to small constants.

\medskip
\noindent\textbf{Update horseshoe prior variance parameters for \(\bm{\beta}_q\).} For each \(q=1,\dots,Q\) and \(r=1,\dots,R\), update the global and local shrinkage parameters as well as the auxiliary variables via the data augmentation scheme of \citet{makalic2015simple}:
\begin{align*}
\rho_{\beta,q,r} &\sim \mathcal{IG}\bigl(1, \frac{1}{\nu_{\beta,q,r}}+\frac{\beta_{q,r}^2}{2\xi_{\beta,q}}\bigr),\\[1mm]
\xi_{\beta,q} &\sim \mathcal{IG}\bigl(\frac{1+R}{2}, \frac{1}{\zeta_{\beta,q}}+\frac{1}{2}\sum_{r=1}^R\frac{\beta_{q,r}^2}{\rho_{\beta,q,r}}\bigr),\\[1mm]
\nu_{\beta,q,r} &\sim \mathcal{IG}\bigl(1, 1+\frac{1}{\rho_{\beta,q,r}}\bigr),\\[1mm]
\zeta_{\beta,q} &\sim \mathcal{IG}\bigl(1, 1+\frac{1}{\xi_{\beta,q}}\bigr).
\end{align*}

\medskip
\noindent\textbf{Update horseshoe prior variance parameters for \(\bm{\delta}\).} Similarly,
\begin{align*}
\rho_{\delta,r} &\sim \mathcal{IG}\bigl(1, \frac{1}{\nu_{\delta,r}}+\frac{\delta_{r}^2}{2\xi_{\delta}}\bigr),\\[1mm]
\xi_{\delta} &\sim \mathcal{IG}\bigl(\frac{1+R}{2}, \frac{1}{\zeta_{\delta}}+\frac{1}{2}\sum_{r=1}^R\frac{\delta_{r}^2}{\rho_{\delta,r}}\bigr),\\[1mm]
\nu_{\delta,r} &\sim \mathcal{IG}\bigl(1, 1+\frac{1}{\rho_{\delta,r}}\bigr),\\[1mm]
\zeta_{\delta} &\sim \mathcal{IG}\bigl(1, 1+\frac{1}{\xi_{\delta}}\bigr).
\end{align*}

\singlespacing
\vspace*{2em}\noindent{\large\MakeUppercase{\textbf{References}}}\vspace*{0.5em}
\printbibliography[heading=none]

\singlespacing
\section{Additional Results, Tables and Figures}

\begin{table}[!ht]
\caption{Summary Statistics of Covariates Included in Regression Specification.}
\begin{center}
\resizebox{\textwidth}{!}{%
\begin{threeparttable}
\begin{tabular}{lrrrrrrrr}
\toprule
  & Mean & Median & SD & Min. & Q25 & Q75 & Max. & Source\\
\midrule
Australia and New Zealand & 0.01 & 0.00 & 0.11 & 0.00 & 0.00 & 0.00 & 1.00 & UN Classification\\
Central Asia & 0.03 & 0.00 & 0.18 & 0.00 & 0.00 & 0.00 & 1.00& UN Classification\\
Eastern Asia & 0.03 & 0.00 & 0.18 & 0.00 & 0.00 & 0.00 & 1.00& UN Classification\\
Eastern Europe & 0.07 & 0.00 & 0.25 & 0.00 & 0.00 & 0.00 & 1.00& UN Classification\\
Latin America and the Caribbean & 0.15 & 0.00 & 0.36 & 0.00 & 0.00 & 0.00 & 1.00& UN Classification\\
Melanesia & 0.01 & 0.00 & 0.10 & 0.00 & 0.00 & 0.00 & 1.00& UN Classification\\
Northern Africa & 0.04 & 0.00 & 0.20 & 0.00 & 0.00 & 0.00 & 1.00& UN Classification\\
Northern America & 0.01 & 0.00 & 0.11 & 0.00 & 0.00 & 0.00 & 1.00& UN Classification\\
Northern Europe & 0.07 & 0.00 & 0.25 & 0.00 & 0.00 & 0.00 & 1.00& UN Classification\\
South-Eastern Asia & 0.06 & 0.00 & 0.24 & 0.00 & 0.00 & 0.00 & 1.00& UN Classification\\
Southern Asia & 0.05 & 0.00 & 0.23 & 0.00 & 0.00 & 0.00 & 1.00& UN Classification\\
Southern Europe & 0.08 & 0.00 & 0.27 & 0.00 & 0.00 & 0.00 & 1.00& UN Classification\\
Sub-Saharan Africa & 0.22 & 0.00 & 0.41 & 0.00 & 0.00 & 0.00 & 1.00& UN Classification\\
Western Asia & 0.11 & 0.00 & 0.32 & 0.00 & 0.00 & 0.00 & 1.00& UN Classification\\
Western Europe & 0.04 & 0.00 & 0.20 & 0.00 & 0.00 & 0.00 & 1.00& UN Classification\\
Female & 0.49 & 0.00 & 0.50 & 0.00 & 0.00 & 1.00 & 1.00& Statistics Austria\\
Contiguity & 0.05 & 0.00 & 0.21 & 0.00 & 0.00 & 0.00 & 1.00& CEPII Gravity Data\\
EU Member & 0.17 & 0.00 & 0.38 & 0.00 & 0.00 & 0.00 & 1.00& European Union\\
Ln Battle Deaths & 2.58 & 0.00 & 3.37 & 0.00 & 0.00 & 5.06 & 11.78 & UCDP GED v23.1\\
Ln Distance of Capitals & 8.17 & 8.48 & 1.04 & 4.09 & 7.43 & 9.02 & 9.82 & CEPII Gravity Data\\
Ln GNIPC & 9.41 & 9.49 & 1.14 & 6.64 & 8.50 & 10.33 & 11.40 & UNDP HDR 2021-22\\
Ln Population & 16.35 & 16.26 & 1.60 & 11.17 & 15.40 & 17.42 & 21.06 & WDI\\
\% Working Age Pop. & 63.76 & 64.89 & 6.45 & 48.56 & 59.77 & 67.90 & 84.58 & WDI\\
Ln Bilateral Migrant Stock & 7.08 & 6.98 & 2.11 & 1.95 & 5.52 & 8.20 & 12.30 & Statistics Austria\\
\bottomrule
\end{tabular}
\begin{tablenotes}
\item \noindent \textit{Note:} SD = standard deviation. Min. = minimum. Max. = maximum. GNIPC = Gross national income per capita. UN = United Nations. CEPII = Centre d'Etudes Prospectives et d'Informations Internationales. UCDP GED = Uppsala Conflict Data Program Georeferenced Event Dataset. UNDP HDR = United Nations Development Programme Human Development Report. WDI = World Development Indicators. Ln is the natural logarithm. GNIPC, Population, the number of battle deaths and \% Working Age Population are averages from 2016 to 2020, while the bilateral migration stock is measured in 2016.
\end{tablenotes}
\end{threeparttable}%
}
\end{center}
\label{table:summarystats}
\end{table}

\begin{figure}[!ht]
    \centering
    \includegraphics[width = \textwidth]{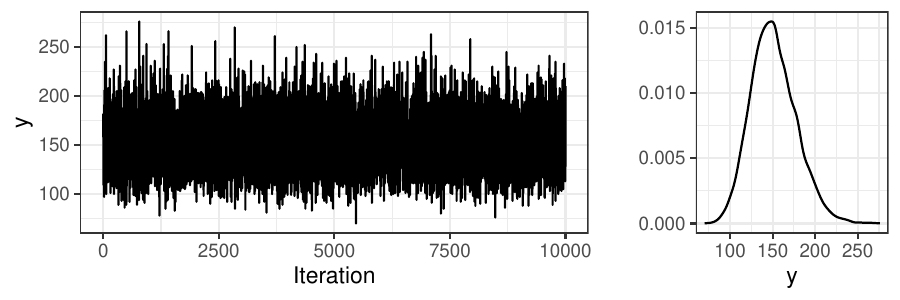}
    \caption{Traceplot of a selected predictive density of a count outcome $y_{i, x}$ (left) and corresponding density estimate (right).}
    \label{fig:convergence}
\end{figure}

\begin{figure}[!ht]
    \centering
    \includegraphics[width = \textwidth]{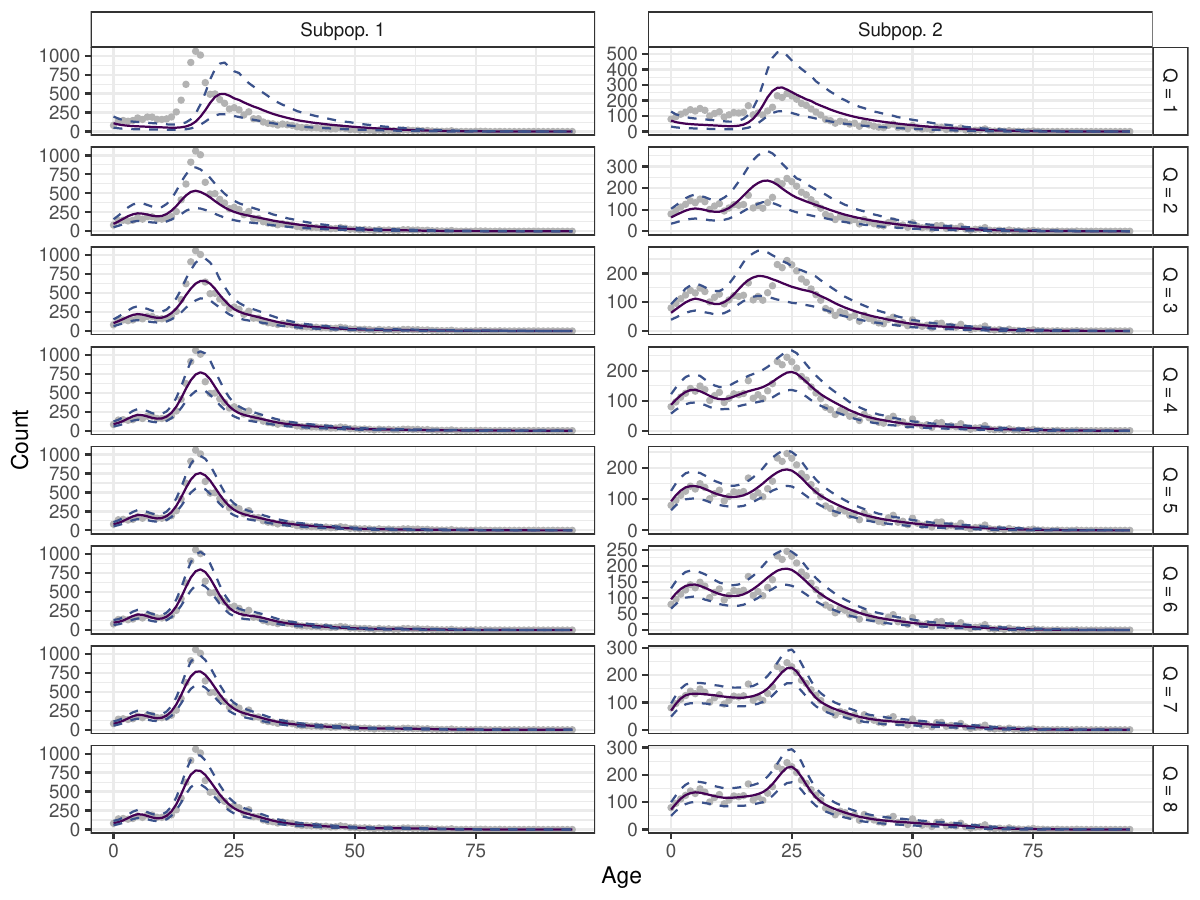}
    \caption{Predicted counts (solid lines) for males (left) and females (right) from Afghanistan as the number of latent factors $Q$ is increased. Points are observed data. Dashed lines are $90\%$ credible intervals.}
    \label{fig:varying_q}
\end{figure}

\begin{figure}[!ht]
    \centering
    \includegraphics[width = \textwidth]{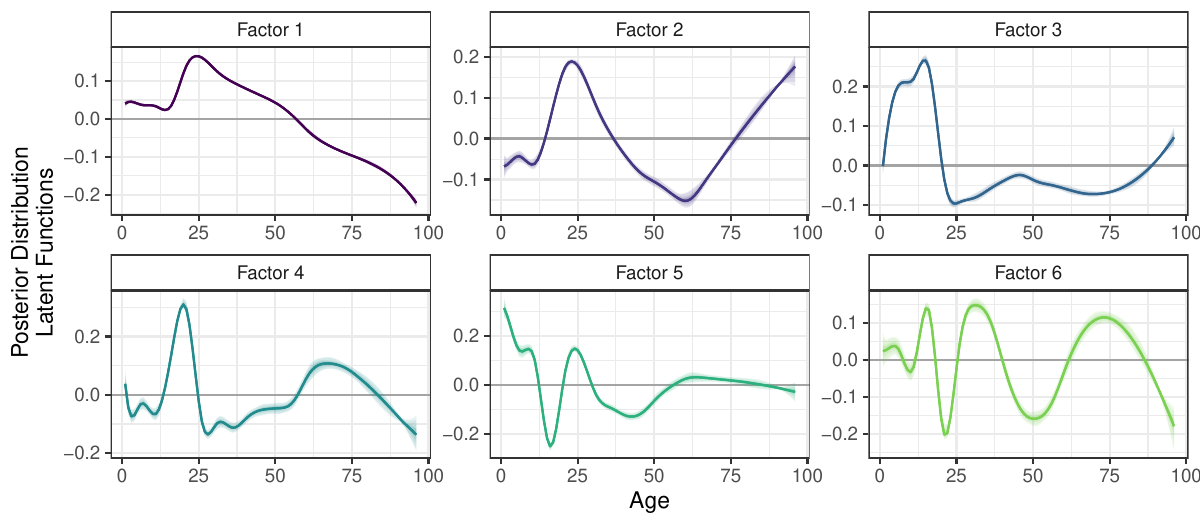}
    \caption{Estimated latent functions $\Phi_q(x)$ as well as 95\% credible intervals.}
    \label{fig:factors}
\end{figure}

\begin{figure}[!ht]
    \centering
    \includegraphics[width = \textwidth]{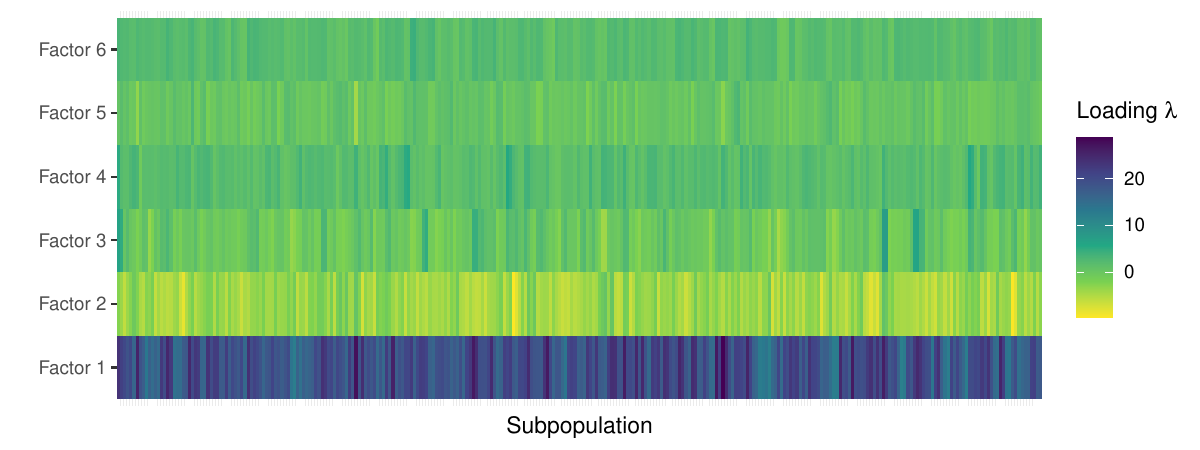}
    \caption{Estimated posterior means of loadings $\lambda_{ix}$ for all 300 subpopulations (columns) and the $Q=6$ estimated factors (rows).}
    \label{fig:loadings}
\end{figure}

\begin{figure}[!ht]
    \centering
    \includegraphics[width = \textwidth]{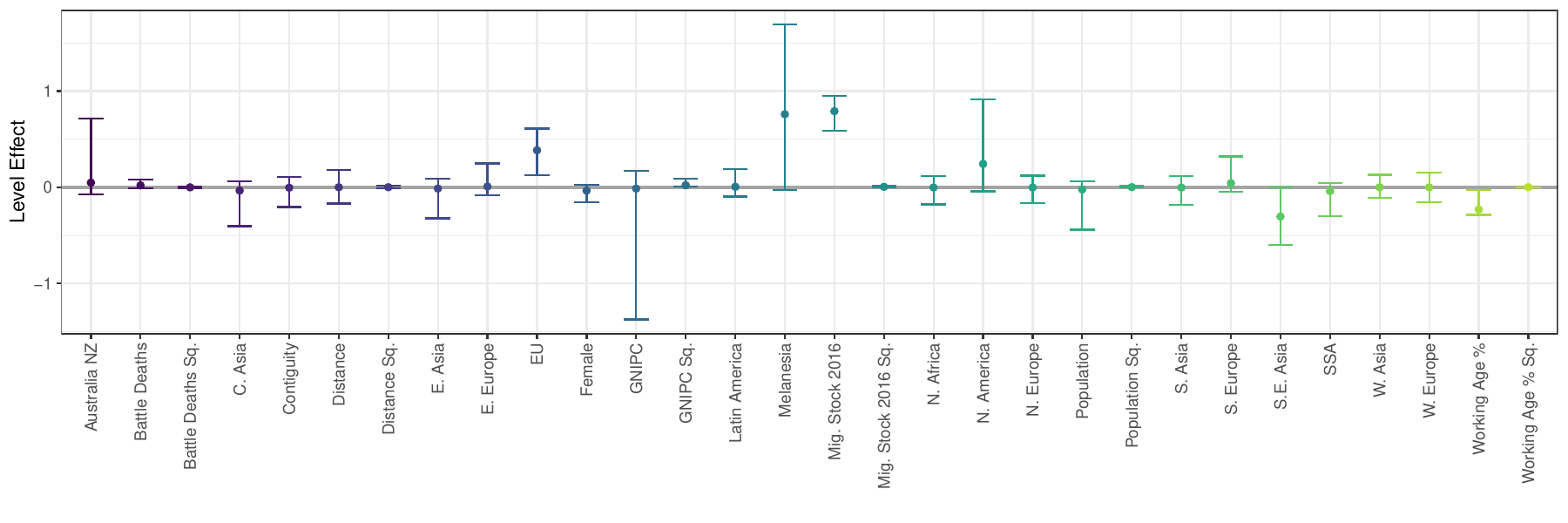}
    \caption{Posterior mean estimates of $\bm{\delta}$ as well as 95\% credible intervals. These estimates describe the average level shift of the demographic process of interest after a unit increase in a given covariate.}
    \label{fig:level_fx}
\end{figure}

\begin{figure}
    \centering
    \includegraphics[width = \textwidth]{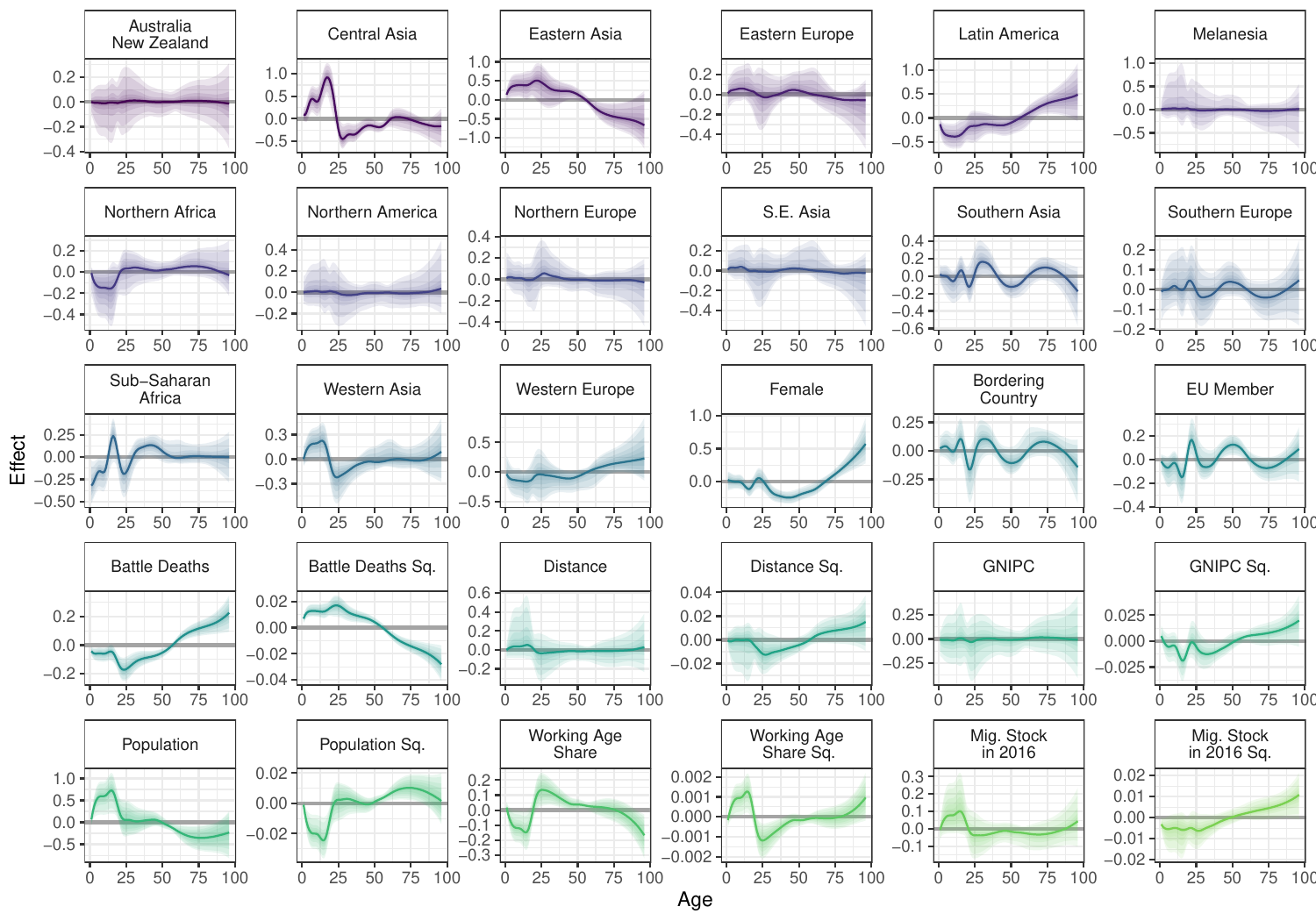}
    \caption{Posterior mean estimates of the shape effects of the included covariates as well as 95\% credible intervals. These estimates describe the average modulation of the demographic process of interest after a unit increase in a given covariate.}
    \label{fig:shape_fx}
\end{figure}

\begin{figure}[!ht]
    \centering
    \includegraphics[width = \textwidth]{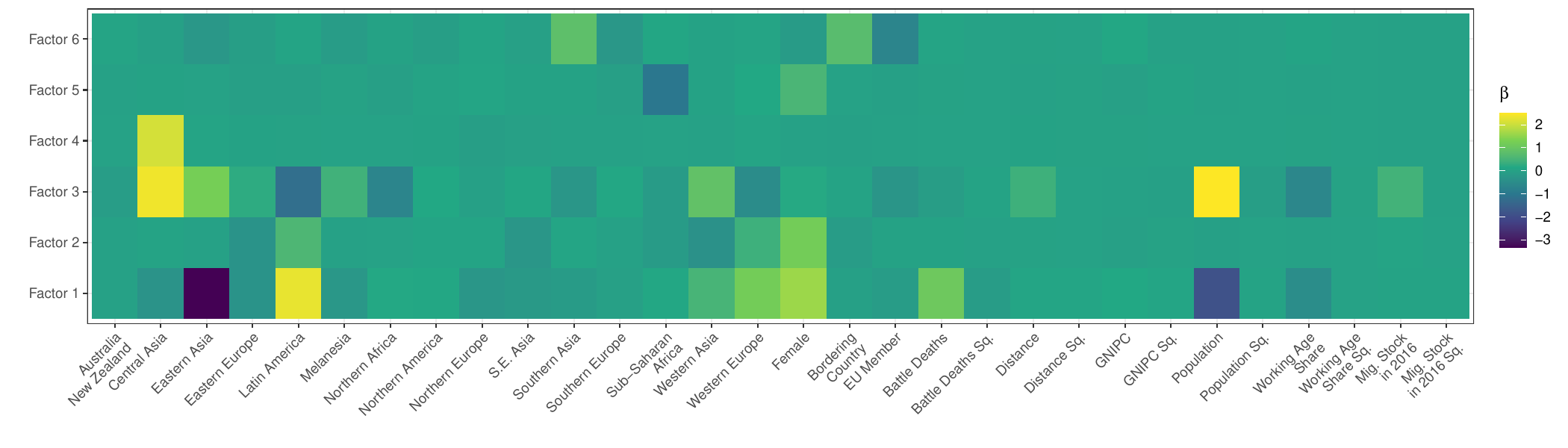}
    \caption{Estimated posterior means for coefficients $\bm{\beta}$ for each covariate / factor combination.}
    \label{fig:beta_factors}
\end{figure}

\begin{figure}[ht]
    \centering
    \includegraphics[width=0.8\textwidth]{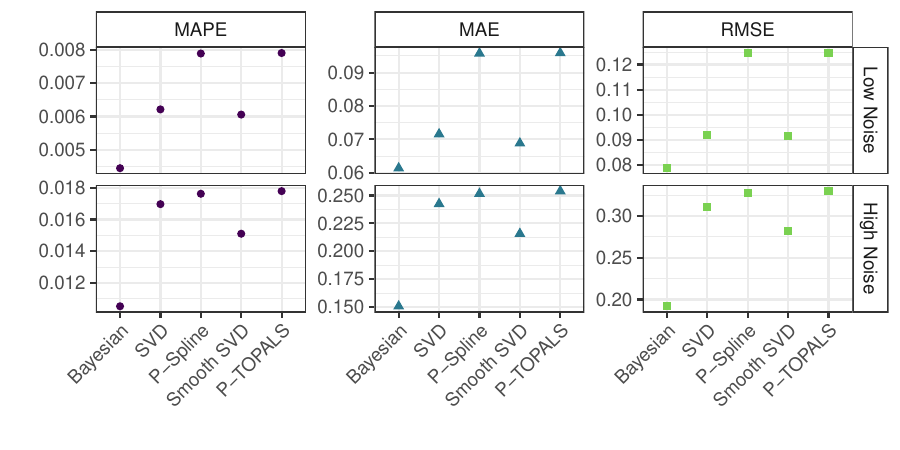}
    \caption{In-sample simulation study evaluation under the baseline simulation scenario. Competing models are shown along the x-axis. MAPE = mean absolute percentage error, MAE = mean absolute error, and RMSE = root mean square error.}
    \label{fig:insample_baseline}
\end{figure}

\begin{figure}[ht]
    \centering
    \includegraphics[width=0.8\textwidth]{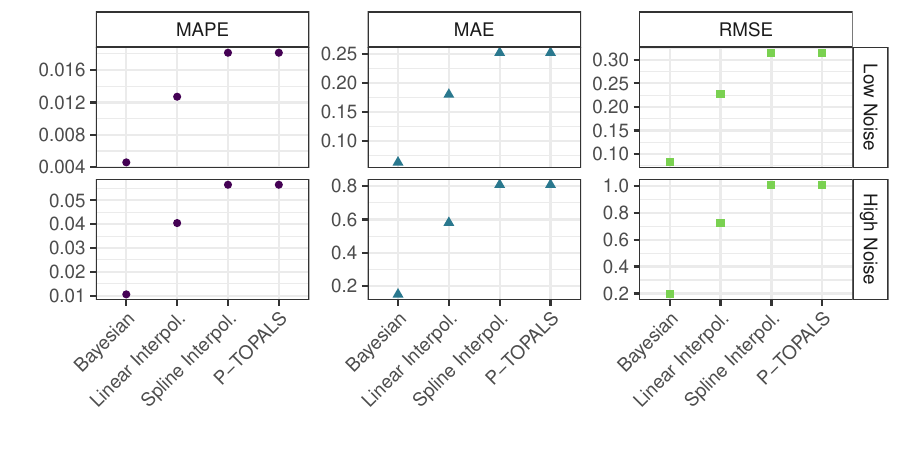}
    \caption{Partially missing data simulation study evaluation. Competing models are shown along the x-axis. MAPE = mean absolute percentage error, MAE = mean absolute error, and RMSE = root mean square error. Metrics are computed only for estimated values of missing observations.}
    \label{fig:missing_data}
\end{figure}

\begin{figure}[ht]
    \centering
    \includegraphics[width=0.8\textwidth]{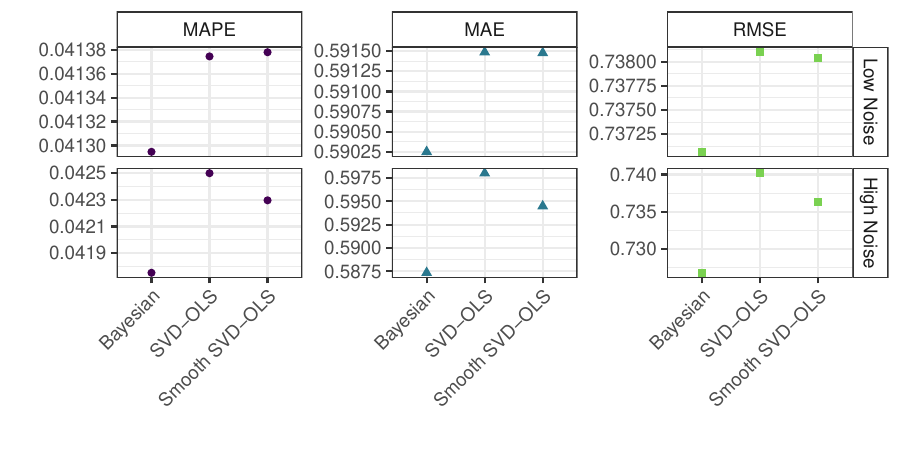}
    \caption{Out-of-sample simulation study evaluation. Competing models are shown along the x-axis. MAPE = mean absolute percentage error, MAE = mean absolute error, and RMSE = root mean square error.}
    \label{fig:oos_evaluation}
\end{figure}

\begin{figure}[ht]
    \centering
    \includegraphics[width=0.8\textwidth]{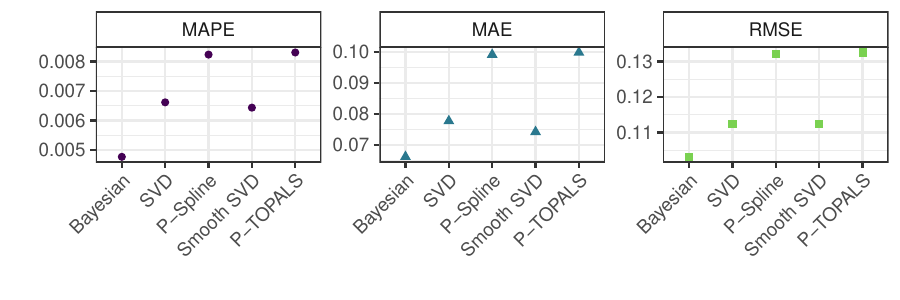}
    \caption{In-sample simulation study evaluation under the outlier simulation scenario. Competing models are shown along the x-axis. MAPE = mean absolute percentage error, MAE = mean absolute error, and RMSE = root mean square error. For this simulation, $\sigma^2=0.1$ is fixed.}
    \label{fig:insample_outliers}
\end{figure}

\begin{figure}[ht]
    \centering
    \includegraphics[width=0.8\textwidth]{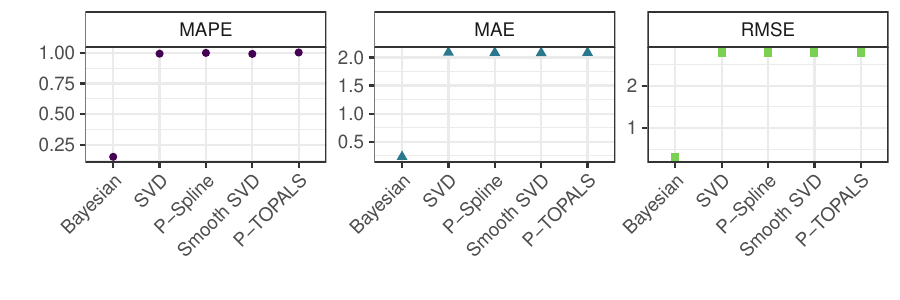}
    \caption{In-sample simulation study evaluation under the sparsity simulation scenario. Competing models are shown along the x-axis. MAPE = mean absolute percentage error, MAE = mean absolute error, and RMSE = root mean square error. For this simulation, $\sigma^2=0.1$ is fixed.}
    \label{fig:insample_sparse}
\end{figure}

\begin{figure}
    \centering
    \includegraphics[width = \textwidth]{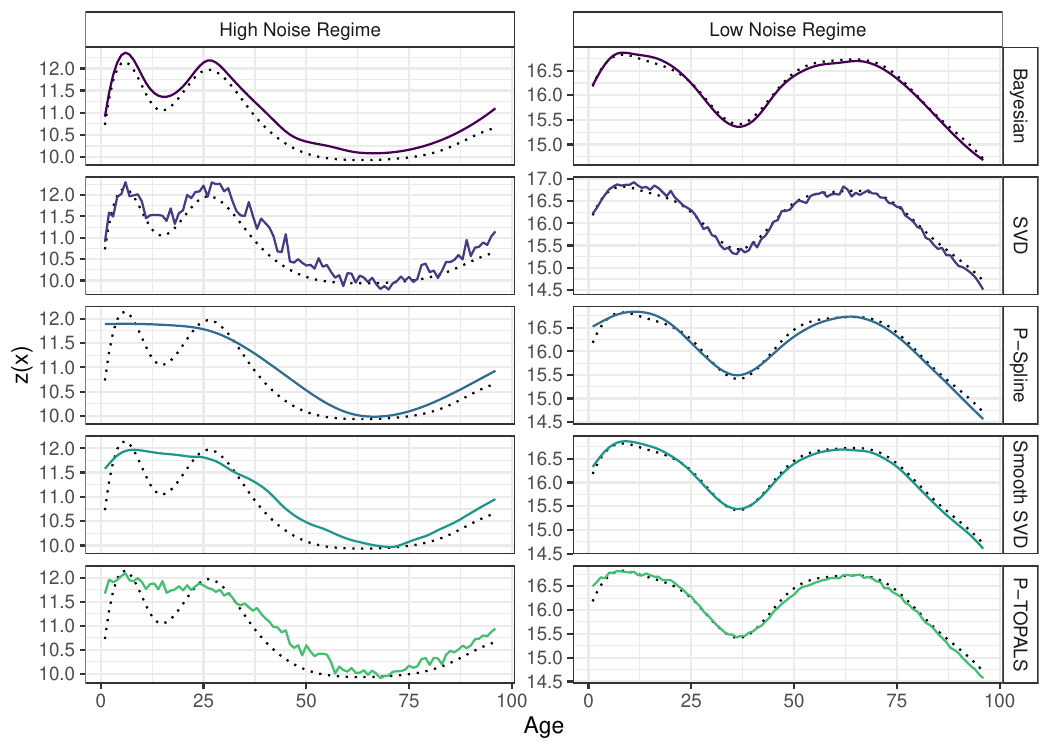}
    \caption{Two selected examples for in-sample fit based on simulated data. Columns correspond to high/low noise regimes. Rows correspond to four different models considered. Refer to the text for details.}
    \label{fig:insample_examples_sim}
\end{figure}

%
%

\end{document}